\documentclass[journal,comsoc]{IEEEtran}
\usepackage{amssymb}
\usepackage{amsfonts}
\usepackage{savesym}
\usepackage{amsfonts,subfigure,multicol,color,verbatim,
graphicx,cite,epsfig,amssymb,amsmath,cases,bm,algorithm,
algorithmic,xcolor,multirow}
\usepackage{pbox}
\usepackage{fontawesome}

\usepackage{pifont}
\usepackage{enumerate}
\usepackage{epstopdf}
\usepackage{tikz}
\usetikzlibrary{matrix,calc,decorations.pathreplacing,decorations.pathmorphing,decorations.fractals,fit,trees,shapes,arrows,positioning,matrix,chains}
\usepackage{bbm}

\usepackage{prettyref}
\usepackage{booktabs}
\usepackage{multirow}
\usepackage{balance}
\usepackage{array}
\usepackage[english]{babel}
\usepackage{caption}
\usepackage{nameref}
\usepackage[hyphens,spaces]{url}
\usepackage{cite}
\usepackage{graphicx}
\graphicspath{{images/}}

\begin{document}

\title{Deep Learning Techniques for Future Intelligent Cross-Media Retrieval}
\author{Sadaqat~ur~Rehman,~\IEEEmembership{Member,~IEEE},  
Muhammad Waqas,~\IEEEmembership{Member,~IEEE},
Shanshan Tu,~\IEEEmembership{Member,~IEEE}, Anis Koubaa, Obaid ur Rehman, Jawad Ahmad,
Muhammad~Hanif,~\IEEEmembership{Member,~IEEE}, 
Zhu Han~\IEEEmembership{Fellow,~IEEE},

\thanks{S. Rehman is an Assistant Professor with the Department of Computer Science, Namal Institute - an associated college of the University of Bradford UK. e-mail: (engr.sidkhan@gmail.com)}
\thanks{S. Tu is with the Faculty of Information Technology, Beijing University of Technology, Beijing China. e-mail: (sstu@bjut.edu.cn)}
\thanks{M. Waqas is with the Faculty of Information Technology, Beijing University of Technology, Beijing China, and also with Department of Computer Science and Engineering, Ghulam Ishaq Khan Institute of Engineering Sciences and Technology, Topi, 23460, Pakistan, e-mail: (engr.waqas2079@gmail.com)}
\thanks{A.~Koubaa is with the Robotics and Internet-of-Things research lab, Department of Computer Science, Prince Sultan University, R\&D Gai-tech Robotics, China and CISTER/INESC TEC and ISEP-IPP, Porto, Portugal.}
\thanks{O.~Rehman is an Assistant Professor in the Department of EE, Sarhad University of Science and IT, Pakistan, e-mail:(obaid.ee@suit.edu.pk)}
\thanks{J.~Ahmad is a Lecturer in the Department of Computer Science, Edinburgh Napier University, UK, email:(jawadkhattak@ieee.org)}
\thanks{M. Hanif is with Department of Computer Science and Engineering,
Ghulam Ishaq Khan Institute of Engineering Sciences and Technology, Topi,
23460, Pakistan, e-mail: (muhammad.hanif@giki.edu.pk)}

\thanks{Z. Han (zhan2@uh.edu) is with the Department of Electrical and Computer
Engineering, University of Houston, Houston, TX 77004, USA.}

\thanks{S. Tu is the corresponding author.}

}

\markboth{}%
{Shell \MakeLowercase{\textit{et al.}}: Bare Demo of IEEEtran.cls for IEEE Communications Society Journals}

\maketitle

\begin{abstract}
With the advancement in technology and the expansion of broadcasting, cross-media retrieval has gained much attention. It plays a significant role in big data applications and consists in searching and finding data from different types of media. In this paper, we provide a novel taxonomy according to the challenges faced by multi-modal deep learning approaches in solving cross-media retrieval, namely: representation, alignment, and translation. These challenges are evaluated on deep learning (DL) based methods, which are categorized into four main groups: 1) unsupervised methods, 2) supervised methods, 3) pairwise based methods, and 4) rank based methods. Then, we present some well-known cross-media datasets used for retrieval, considering the importance of these datasets in the context in of deep learning based cross-media retrieval approaches. Moreover, we also present an extensive review of the state-of-the-art problems and its corresponding solutions for encouraging deep learning in cross-media retrieval. The fundamental objective of this work is to exploit Deep Neural Networks (DNNs) for bridging the ``media gap'', and provide researchers and developers with a better understanding of the underlying problems and the potential solutions of deep learning assisted cross-media retrieval. To the best of our knowledge, this is the first comprehensive survey to address cross-media retrieval under deep learning methods.
\end{abstract}

\section{Introduction}
\IEEEPARstart{S}{ocial} media websites (e.g., Facebook, Youtube, Instagram, Flickr, and Twitter) have tremendously increased the volume of multimedia data over the Internet. Consequently, considering this large volume of data and the heterogeneity of the data sources, data retrieval becomes more and more challenging. Generally, multimodal data (i.e., data from sources, e.g., video, audio, text, images) are used to describe the same events or occasions. For instance, a web page describes similar contents of an event in different modalities (image, audio, video, and text). Therefore, with a large amount of multimodal data, the accurate result of a search concerning the information of interest decreases. The evolution of different search algorithms for indexing and searching multimodal data contributed positively to searching for information of interest efficiently. Nevertheless, they only work in a single-modality-based search, comprising two main classes: content-based retrieval and keyword-based retrieval \cite{gasser2019towards}.

In the last few years, many cross-media retrieval methods have been proposed \cite{rehman2018benchmark, peng2018overview, dong2018semi, xia2018cross, liu2018multi, shu2018crossfire, xu2018deep }. However, Canonical Correlation Analysis (CCA) \cite{hardoon2004canonical} and Partial Least Square (PLS) \cite{rosipal2005overview, sharma2011bypassing} are usually adopted to explicitly project different modality data to a common space for similarity measurement. In the Bilinear Model (BLM) \cite {tenenbaum2000separating}, different modality (e.g., text and image) data are projected to the same coordinates as it learns a common subspace. Generalized Multiview Analysis (GMA) \cite{sharma2012generalized} can be used to combine CCA, BLM, and PLS for solving cross-media retrieval task. Gong \textit{et. al.} \cite{gong2014multi} proposed a variant CCA model by incorporating the high-level semantic information as a third view. Ranjan et al. \cite{ranjan2015multi} also introduced a variant of CCA called multilabel Canonical Correlation Analysis (ml-CCA) for learning the weights of shared subspaces using high-level semantics called multi label annotations. Rasiwasia \textit{et al.} \cite{rasiwasia2014cluster} proposed a cluster CCA method to learn discriminant isomorphic representations that maximize the correlation between two modalities while distinguishing the different categories. Sharma \textit{et. al.} \cite{sharma2012generalized} proposed a variant of Marginal Fisher Analysis (MFA) called Generalized Multiview Marginal Fisher Analysis (GMMFA).

\begin{table*}
\begin{center}
\caption{Comparison of existing survey articles on deep learning and cross-media retrieval. \ding{52} represents that the topic is covered, \ding{56} represents the topic is not covered, and \ding{106} represents the topic is partially covered.  }
    \label{survey-summ}

\begin{tabular}{|m{0.5cm}|m{0.4cm}|m{4.4cm}|c|c|c|c|c|c|c|}
\hline
\center{\textbf{Ref.}} & \center{\textbf{ Year}} & \center{\textbf{Topic}} &\multicolumn{4}{c|}{\textbf{Deep Learning}} &  %
    \multicolumn{3}{c|}{\textbf{Cross-media Retrieval}} \\
\cline{4-10}

 &  &  & Supervised & Unsupervised& Pairwise&Rank & Representation & Alignment&Transalation \\
\hline
 \cite{schmidhuber2015deep} &2015 & Deep learning in neural networks: An overview& \ding{52}& \ding{52}&  \ding{52}& \ding{52}&\ding{56}&\ding{56}&\ding{56} \\
\hline
\cite{lecun2015deep} & 2015&Deep Learning &\ding{52} & \ding{52}&\ding{52} &\ding{52}&\ding{56}&\ding{56}&\ding{56} \\
\hline
\cite{liu2017survey}&2017 &A survey of deep neural network architectures and their applications. & \ding{52}& \ding{52}&\ding{52} &\ding{52}&\ding{56}&\ding{56}&\ding{56} \\
\hline
\cite{ahmad2019deep} &2019 & Deep learning: methods and applications& \ding{52}& \ding{52}&\ding{52} &\ding{52}&\ding{56}&\ding{56}&\ding{56} \\
\hline
\cite{deng2014tutorial} &2014 & A tutorial survey of architectures, algorithms, and applications for deep learning&\ding{52} &\ding{52} & \ding{52}&\ding{52}&\ding{56}&\ding{56}&\ding{56} \\
\hline
\cite{pouyanfar2018survey} & 2018&A survey on deep learning: Algorithms, techniques, and applications & \ding{52}&\ding{52} &\ding{52} &\ding{52}&\ding{56}&\ding{56}&\ding{56} \\
\hline
\cite{arulkumaran2017deep} &2017 & Deep reinforcement learning: A brief survey&\ding{52} &\ding{52} &\ding{56} &\ding{56}&\ding{56}&\ding{56}&\ding{56} \\
\hline
\cite{hussein2017imitation} & 2017& Imitation learning: A survey of learning methods& \ding{52}&\ding{52} &\ding{52} &\ding{106}&\ding{56}&\ding{56}&\ding{56} \\
\hline
\cite{chen2014big} & 2014& Big data deep learning: challenges and perspectives&\ding{52} &\ding{52} & \ding{56}&\ding{56}&\ding{56}&\ding{56}&\ding{56} \\
\hline
 \cite{najafabadi2015deep}& 2015&Deep learning applications and challenges in big data analytics &\ding{52}&\ding{52} &\ding{56} &\ding{56}&\ding{56}&\ding{56}& \ding{56}\\
\hline
\cite{hordri2017systematic} &2017 &A systematic literature review on features of deep learning in big data analytics &\ding{52} & \ding{52}& \ding{52}&\ding{56}&\ding{56}&\ding{56}&\ding{56} \\
\hline
\cite{peng2017overview} & 2017&An overview of cross-media retrieval: Concepts, methodologies, benchmarks, and challenges &\ding{52} & \ding{52}&\ding{56} &\ding{106}&\ding{52}&\ding{56}&\ding{56} \\
\hline
\cite{wang2016comprehensive} & 2016&A comprehensive survey on cross-modal retrieval &\ding{52} & \ding{52}&\ding{52} &\ding{52}&\ding{52}&\ding{56}&\ding{56} \\
\hline
\cite{liu2010cross} & 2010&Cross-media retrieval: state-of-the-art and open issues& \ding{52}& \ding{52}& \ding{56}&\ding{56}&\ding{52}&\ding{56}&\ding{56} \\
\hline
\textbf{Our work} & 2020 &Deep Learning Techniques: Evolving Machine
Intelligence for Future Intelligent Cross-media
Retrieval &\ding{52}  &\ding{52}  &\ding{52}  &\ding{52} &\ding{52} &\ding{52} & \ding{52} \\
\hline
\end{tabular}
\end{center}
\end{table*}

Even though every aforementioned contribution provide vital contribution in cross-media retrieval society, still these methods lack satisfactory performance. The key reason is that conventional feature learning techniques hardly tackle the problem of image understanding, but visual features representation between images and text is highly dependent on cross-media retrieval.  Recently, deep learning models have made significant development in fields such as computer vision \cite{tu2017csfl, benjdira2019unsupervised}, engineering \cite{rehman2016face}, health \cite{yang2018clinical} and hydrology \cite{Rehman2019water}. Donahue \textit {et. al.} \cite{donahue2014decaf} proposed a deep eight-layer neural network called DeCAF, which confirmed that Convolution Neural Network (CNN) features are helpful for various feature extraction tasks.

 In this paper, we investigate different deep learning approaches applied in the domain of cross-media search, which are indispensable for the adoption and implementation of cross-media retrieval. DNN is designed to simulate the neuronal structure of the human brain, and represents a powerful approach to naturally deal with the correlations of multi media. For this purpose, several researchers have explored DNNs for using it in the search and retrieval of data from heterogenous sources. Although, the latest research in the field of DNN-based methods for cross-media retrieval has achieved better performance \cite{peng2016cross}, however, there are still significant improvements needed in this area. 
 
We explore the following three main challenges for using deep learning techniques in cross-media retrieval.

 \begin{figure}
\includegraphics[width=10.5cm, height=10cm]{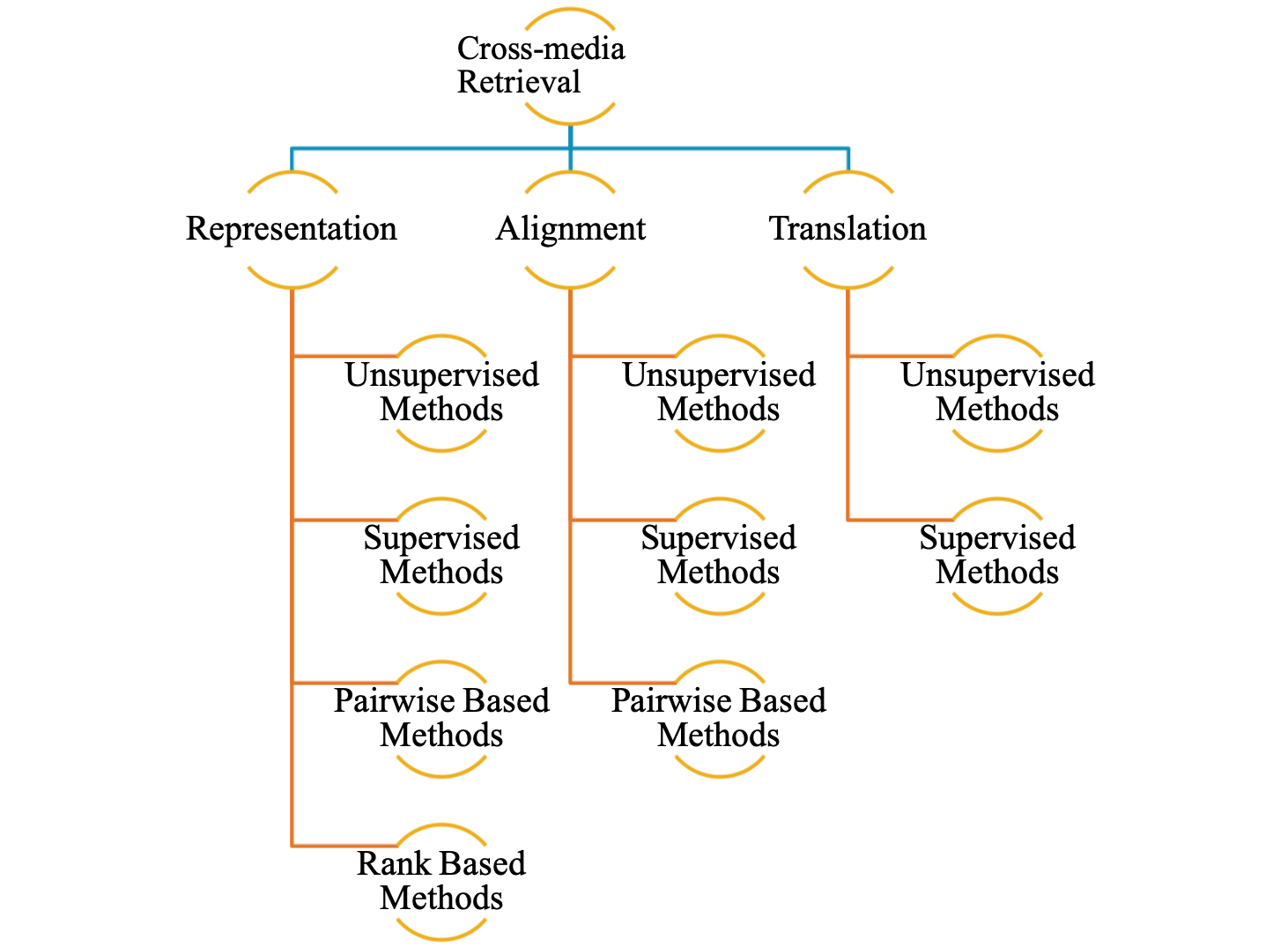}
\caption{Taxonomy of the proposed work.}
\label{taxonomy}
\end{figure}

\begin{enumerate}
\item \textbf{Representation.} It aims to learn the representation of cross-media data in an optimal way to mitigate its redundancy. This is a challenging task in cross-media retrieval since data is heterogeneous. For instance, the text is normally symbolic while audio and video modalities are represented as signals. Therefore, learning the representation of individual modality in a common semantic space is a challenging task.
  
\item \textbf{Alignment.} In this procedure, the key objective is to find the correlation between elements from cross modalities to mitigate the modality-to-modality mismatch issue. For instance, we want to align each human action image into a video showing a series of different human actions. To achieve this, we need to measure the similarity distance between different modalities and deal with other correlation uncertainties.

\item \textbf{Translation.} It shows the correlation mapping of data across different modalities, since data is heterogeneous and the relationship between cross modalities is hard to identify. For instance, an image can be described in various different ways, and a single perfect translation may not exist. Therefore, it is hard to choose an appropriate translation for a particular task, where multiple parameters are crucial. Particularly, there is no appropriate correct answer to a query in translation. As there is no common concept of translation to chose which answer is right and which is wrong.
\end {enumerate}

For each of the aforementioned problems in cross-media retrieval, we provide a taxonomy of classes and sub-classes. A detailed taxonomy is provided in Fig. \ref{taxonomy}. We found out that some key issues of deep learning in cross-media retrieval on concepts, methodologies and benchmarks are still not clear in the literature. To tackle the aforementioned challenges, we investigate the DNN-based methods assisted cross-media retrieval.

\subsection{Comparison with Related Surveys Article}

Our current survey article is unique in a sense that it
comprehensively covers the area of DNNs-based cross-media retrieval. There is
no prior detailed survey article that jointly considers DNNs and cross-media retrieval, to the best of our knowledge. Though there is an extensive literature on survey articles on DNNs or cross-media retrieval, but these survey
articles either focus on DNNs or cross-media retrieval, individually.

General surveys regarding deep learning are discussed in \cite{lecun2015deep, liu2017survey, ahmad2019deep, pouyanfar2018survey}. Surveys dealing with only cross-media retrieval domain are presented in \cite{liu2010cross}. Our work is closely related to \cite{wang2016comprehensive, peng2017overview}; however, they cover the broader picture of cross-media retrieval domain whereas, our work is more focus on DL-based cross-media retrieval. Furthermore, we provide a novel taxonomy according to the challenges faced by multi-modal deep learning approaches in solving cross-media retrieval, namely: representation, alignment, and translation. To the best of our knowledge, this is the only work till date, which provide a detail survey of DL-based methods in solving cross-media retrieval challenges (representation, alignment and translation). A summarized comparison of survey articles on DL and cross-media retrieval are provided in Table \ref{survey-summ}. 

\subsection{Our Contributions}

To summarize, our main objectives in this paper are as follows.

\begin{figure}
\center
\includegraphics[width=8.5cm,height=7.5cm]{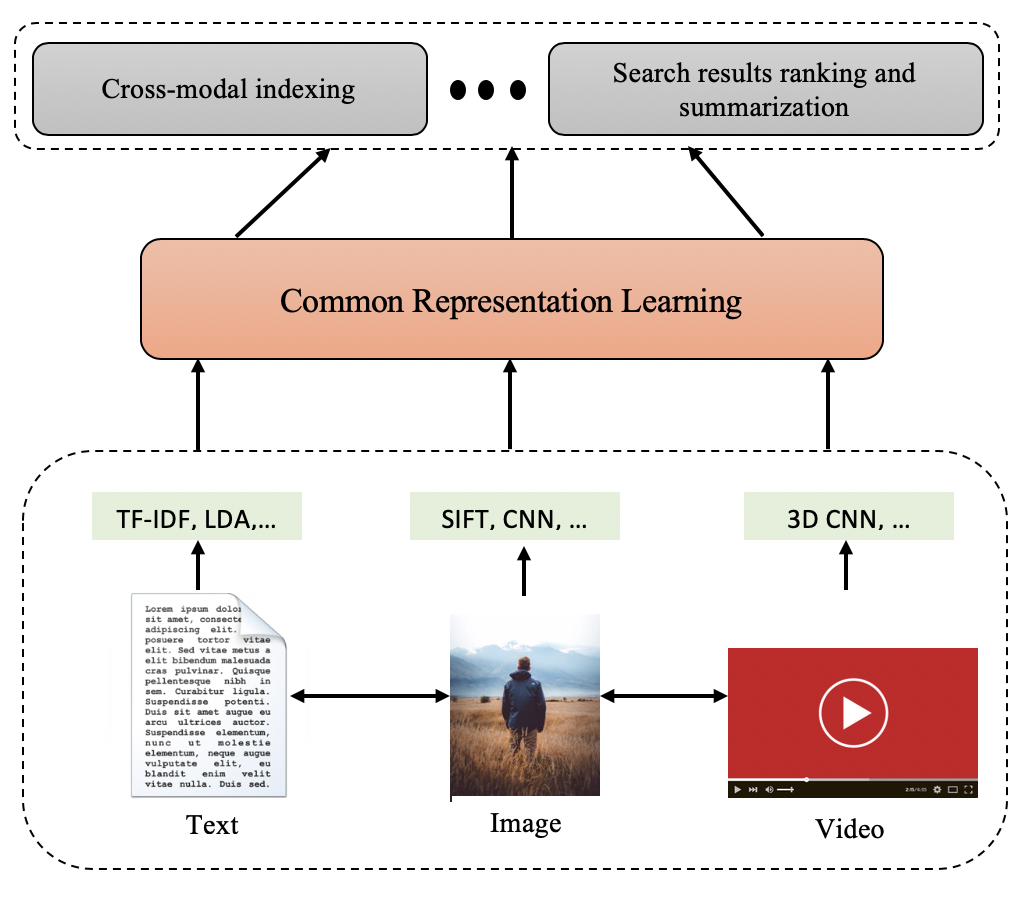}
\caption{A generalized framework of cross-media retrieval system.}
\vspace{-5mm}
\label{fig:gen}
\end{figure}

\begin{itemize}
\item Provide an up to date survey on the current advancement
in cross-modal retrieval. This provides an added value as compared to previous surveys, which represents substantial benefits for understanding the trends in cross-media retrieval rapidly. 
\item  Provide a useful categorization of cross-media retrieval under DNN approaches. The contrasts between various types of techniques 
are expounded, which are helpful for readers to better understand various deep learning techniques used in cross-media retrieval.
\item A detailed explanation of almost every cross-media dataset is provided. Furthermore, its advantages and disadvantages are also discussed to facilitate the developers and researchers choosing a better dataset for their learning algorithms. 
\item Present the key challenges and opportunities in the area of cross-media retrieval and discuss open future research challenges.
\end{itemize}

\section{An Overview of Cross-media Retrieval and Deep Learning}
Before probing in to the depth of this paper, we want to initiate with the fundamental concepts of cross-media retrieval and deep learning techniques. We divide this section into diverse subsection such as, cross-media retrieval is discussed in subsection~A. Moreover, the deep learning techniques in subsection~B to discuss different algorithms for representations. Finally, the subsection~C explain why DL is important for cross-media retrieval?

\subsection{Cross-media Retrieval}
Cross-media retrieval represents the search for different modalities (e.g., images, texts, videos) by giving any individual modality as an input. The generic framework of cross-media retrieval is shown in Fig. \ref{fig:gen}, in which data is represented in different modalities such as text, image, and video. Different algorithms (e.g., CNN, SIFT, LDA, TF-IDF, etc.) are applied to learn the feature vectors of individual modality. Furthermore, in the case of joint semantic space for multimodal data, cross-media correlation learning is performed for feature extraction. Finally, the semantic representations allow the cross-media retrieval to perform search results ranking and summarization. 

It is important to note that cross-media retrieval is different from other correlation matching approaches between various media types (image and text). For example, correlation matching approaches \cite{yang2017image, vinyals2015show} are used to generate the text descriptions of image/video only, whereas the cross-media retrieval approach endeavor to retrieve text from different modalities data image/video and vice versa. Methods of image annotation \cite{ballan2014cross} are used to assign most relevant tags to images for descriptions, whereas in cross-media retrieval, the text also represents sentences and paragraph descriptions instead of only tags.

Cross-media retrieval is an open research issue in real-world applications. With the popularity of social media platforms (i.e., Facebook, Twitter, Youtube, Flickr and Instagram) different types of media (images, videos, texts) are flooding over the Internet. To tackle this issue, different cross-media retrieval approaches have been proposed \cite{8695043, 8643797, 8673892, 8691806, 8716706}. However, in this paper we only consider DNNs-based cross-media retrieval approaches for information utilization to learn the common representations. As, DNNs-based approaches leverage the performance of different learning algorithms in cross-media retrieval domain. Moreover, to our knowledge this is the only survey mutually consider DNNs and cross-media retrieval. We categorize the DNN-based methods for the individual challenge of cross-media retrieval into four classes: (1) unsupervised methods, (2) supervised methods, (3) pairwise based methods, and (4) rank based methods. 

\begin{enumerate}

\item \textbf{Unsupervised methods.} Unsupervised methods leverage co-occurrence information instead of label information to learn common representations across data with different modalities. Specifically, these methods treated different modalities of data existing in a common multi-modal document as the same semantic. For instance, a website page contains both text and pictures for the outline of same theme. Specifically, users get information from both images or texts to get idea of a particular event or topic in a webpage. 

\item \textbf{Supervised methods.} In supervised methods, label information is used to learn common representations. These methods increase the correlation among intra-class samples and decrease the correlation among inter-class samples to obtain good discriminating representations. However, getting annotated data is costly and laborious because of manual labelling.

\item \textbf{Pairwise based methods.} These methods are used to learn common representations through similar/dissimilar pairs, in which, a semantic metric distance is learned between data of various modalities. 

\item \textbf{Rank based methods.} These methods are used to learn common representations for cross-media retrieval through learning to rank. 
\end{enumerate}

\subsection{Deep Learning Techniques}
Deep Learning (DL) is a sub-class of Machine Learning (ML). DL networks are a kind of neural network that discovers important object features. These algorithms attempt to learn (multiple levels of) representation
by using a hierarchy of multiple layers. If the system is provided with a large amount of
information, it begins to understand it through feature extraction and respond in useful
ways. Most of the deep learning algorithms are built on neural network architectures, due
to this reason they are often called as Deep Neural Networks (DNN).

Different DL architectures (Deep Neural Network, Convolution Neural Network,
Deep Belief Networks, Recurrent Neural Network) are successful in solving many computer vision problems efficiently, where the solutions are difficult to obtain analytically.
These problems include handwritten digit recognition, optical character recognition, object classification, face detection, Image captioning and facial expression analysis \cite{bengio2013representation, schmidhuber2015deep, lecun2015deep}.

Currently, DL algorithms are also tested in interdisciplinary research domains, such
as bio-informatics, drug design, medical image analysis, material inspection, agriculture
and hydrology \cite{Rehman2019water, bovsnavcki2019deep, stephenson2019survey, bastian2019visual, alhnaity2019using}. The processing and evolution of these fields are dependent on deep
learning, which is still evolving and in need of creative ideas \cite{cirecsan2012multi, krizhevsky2012imagenet, marblestone2016toward}. 

\subsubsection{Evolution and Classification of Deep Learning Techniques}

\begin{figure}
  \centering
  \includegraphics[width=0.5\textwidth]{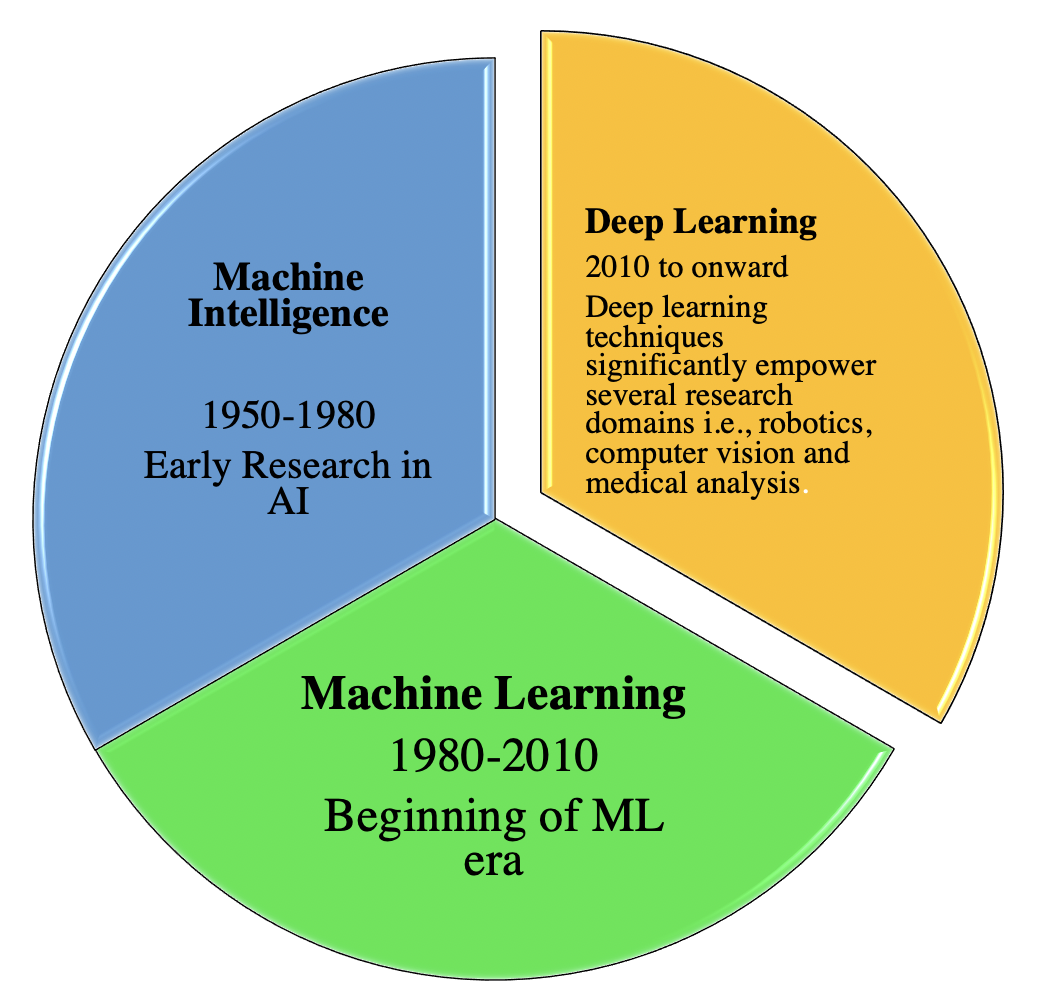}
  \caption{An overview of the evolution of deep learning from conventional Machine Intelligence and Machine Learning paradigms.}
  \label{his-dl}
\end{figure}

Since the early excitement stirred by ML in the 1950s,
smaller subsets of machine intelligence have been impacting
a myriad of applications over the last three decades as
shown in Fig. \ref{his-dl}. Initially, the term \textit{``deep learning''} was presented to the community of machine learning by Rina Dechter in 1986 \cite{AAA12, lecun2015deep}, and Igor Aizenberg and his colleagues to artificial neural networks in 2000, in boolean threshold neurons domain \cite{AAA14, AAA15}. In 1965, Alexey Ivakhnenko and Lapa published the primary general learning algorithm for feed-forward, supervised, multi-layer perceptrons \cite{A91}. Moving forward in 1980, Kunihiko Fukushima introduced Neocognitron in computer vision domain \cite{A92}. Furthermore, Yann LeCun applied standard backpropagation algorithm to deep neural network for handwritten recognition in 1989 \cite{A93, A94, A95, A96}. 

Although, deep learning has existed for more than three decades however, they have recently gain interest in the machine learning community. Before 2006, the deep learning method was a complete failure in training large deep architectures. In 2006, the revolution to successful training schemes for deep architectures originated with the algorithms for training Deep Belief Networks (DBNs) by Hinton \textit{et al.} \cite{A97} and autoencoders by  Ranzato et al. \cite{A98} and Bengio \textit{et al.} \cite{A99} based on unsupervised pre-training followed by supervised fine-tuning. Following the same path, different approaches were proposed to deal with the aforementioned issues under different circumstances. 

 Before 2011, CNNs did not succeed in efficiently solving computer vision problems. However, in 2011, CNNs achieved superhuman performance in a visual pattern recognition contest. In 2012, the success of deep learning algorithms in image and object recognition were started. However, backpropagation algorithm had been used for decades to train CNNs, and Graphical Processing Unit (GPU) implementations of Neural Networks (NNs) for years, comprising CNNs \cite{A100, A101}.Moreover, in the same year CNNs also won ICDAR Chinese handwriting contest. In May 2012, CNNs won ISBI image segmentation contest \cite{A102}, which significantly attracted researcher’s attention. Ciresan \textit{et al.} showed how max-pooling CNNs on GPU can affectedly enhance several computer vision benchmark records at CVPR 2012 \cite{A103}. Following the same path, in October 2012, Krizhevsky \textit{et al.} \cite{krizhevsky2012imagenet} showed the dominancy of DNNs over shallow machine learning methods by winning the large-scale ImageNet competition over a large margin. 

Researchers believe that the victory of ImageNet in Large Scale Visual Recognition Challenge (ILSVRC) 2012 anchored the begin of \textit{``deep learning revolution''} that has revolutionize the Artificial Intelligence (AI) industry \cite{A104}.

\subsection{Why DL for Cross-media Retrieval?}
Before going in detail, it is useful to understand the reason of applying DNNs to cross-media retrieval. There are several DNNs attractive characteristics that make it unique such as (1) end-to-end learning model, (2) efficiency boost up using back-propagation training, and (3) the performance of DNNs increase as the size of data increase \cite{moreira2019contextual, zhang2019deeprec, you2019hierarchical}. Furthermore, the architecture of DNNs are hierarchal and trained end-to-end. The main advantage using such architecture is when dealing multimedia data. For example, a webpage contains textual data (reviews \cite{zheng2017joint}, tweets \cite{gong2016hashtag}), visual data (posts, scenery images), audio data and video data. Here modality-specific features extraction will be complex and time consuming. Suppose, if we have to process textual data, initially we need to perform expensive and time consuming pre-processing (e.g., keywords extraction, main topic selection). However, DNNs have the ability to process all the textual information in a sequential end-to-end manner \cite{zheng2017joint}. Therefore, these advances in the architecture of DNNs make it very suitable for multi-modal tasks \cite{zhang2017bjoint} and we urge for indispensable neural end-to-end learning models. 

As for as the interaction-only settings (i.e. matrix completion) are concerned, DNNs are necessary in dealing huge number of training data and gigantic complexity. He \textit{et al.} \cite{he2017neural} overcome the performance gain of conventional Matrix Factorization (MF) method by using Multi Layer Perceptron (MLP) to approximate the interaction function. Moreover, typical ML models (i.e., BPR and MF) also achieve best performance on interaction-only data when trained with momentum-based gradient descent \cite{tay2018latent}. Nevertheless, these models also take the benefit of current DNNs based improvements such as Batch normalization, Adam, and optimize weight initialization \cite{he2017neural, zhang2018neurec}. It is fact that most of the Cross-media retrieval algorithms have adopted DNNs-based structure to improve its performance such as Deep Canonical Correlation Analysis (DCCA) \cite{liu2019new}, Deep Canonically Correlated Auto-Encoder (DCCAE) \cite{wang2016deep}, and Discriminative Deep Canonical Correlation Analysis (DisDCCA) \cite{elmadany2016multiview}. Therefore, DL is significantly useful tool for today's research and industrial environment for the advancement of cross-media retrieval methods. 

We summarize some of the useful strengths of DNNs based cross-media retrieval models, which are as follows:

\subsubsection {Flexibility}

The DNNs based approaches are also known as global learning due to its vast application domain. Currently, the flexibility of DL methods further boost up with the invent of well-known DL frameworks i.e., Caffe, Tensorflow, Pytorch, Keras, Theano, and MXnet. Each of the aforementioned framework has active community and support. This make development and engineering efficient and easier. For instance, concatenation of different neural models become easier, and produce more powerful hybrid structures. Hence, the implementation of hybrid cross-media retrieval models become easier to capture better features and perform well.  

\subsubsection {Generalization}
This property of DL methods make it very demanding and unique. It can be used in many different applications and with different data types. For example, in the case of transfer learning the DL-based method have the ability to share knowledge across different tasks. As, DL algorithms capture both low and high level features, they may be beneficial to perform other tasks \cite{bengio2013representation}. Andreas \textit{et al.} \cite{andreas2015deep} and Perera \textit{et al.} \cite{perera2019deep} showed the successful performance of DNNs-based methods in transfer learning.

\subsubsection{Nonlinear Transformation}

DNNs based models have the ability to process the non-linearity in data using non-linear activation functions i.e., \textit{sigmoid, relu} and \textit{tanh}. This helps the models to capture complex patterns within the dataset. Traditional cross-media retrieval methods such as CCA, BLM and Linear Discriminant Analysis (LDA) are linear models, which need DNNs-based methods to retrieve nonlinear features. For example, in DCCA, initially DNNs are used to extracts nonlinear features and then uses linear CCA to calculate the
canonical matrices. It is well-know that neural networks have the ability to approximate any continuous function by fluctuating the activation functions \cite{abc}.

\subsubsection{Robust}

DL based methods do not need manually feature extraction algorithms rather feature are learned in an end-to-end manner. Hence, the system achieve better performance despite the variations of the input data. The authors of \cite{Gaurav} and \cite{Wicker2019RobustnessO3} showed the robustness of DL against adversarial attacks in visual recognition application.

\section{Cross-media Datasets}
Dataset plays a critical role in the evaluation of learning algorithm. Its selection is very important for feature extraction and training of different DL algorithms. We summarized some of the well-known cross-media datasets below, and Table \ref{tab:dataset} depicts a comparison evaluation among them.

\begin{enumerate}
\item \textbf{Wikipedia:} this dataset is largely used in cross-media domain to evaluate the performance of different learning algorithms. The dataset consists of 2866 image-text pairs of 10 distinct classes accumulated from Wikipedia's articles.  

\item \textbf{NUS WIDE:} A popular dataset in cross-media community after Wikipedia dataset. This dataset contains 269,648 labeled images of 81 different concepts from Flickr. Every image in the dataset is aligned with associated user tags called image-text pair. Overall, the dataset contains 425,059 unique tags that are associated with these images. Nevertheless, to enhance the quality of tags, those tags were pruned that appear less than 100 times and do not exist in WordNet \cite{miller1995wordnet}. Hence, 5,018 unique tags are included in this dataset.

\item \textbf{Pascal VOC:} the dataset consists of 20 distinct classes of image-tag pairs having 5011 training pairs and 4952 testing pairs. Although, some images are labeled more than twice. However, in the literature some studies have selected uni-labelled images, which results in 2808 and 2841 training and testing pairs, respectively \cite{sharma2012generalized}. The image feature chosen were GIST and color \cite{hwang2012reading}, and histogram whereas; text features were 399-dimensional tag occurrence.

\item \textbf{FB5K:} The dataset contains 5,130 image-tag pairs with associated users' feelings, which is accumulated from Facebook \cite{ur2018facebook5k}. Furthermore, this dataset is categorized into 80\% and 20\% for training and testing image-text pairs. 

\item \textbf{Twitter100K:} This dataset is made up of 100,000 image-text pairs collected from Twitter. It exploited 50,000 and 40,000 image-text pairs for training and testing respectively. Moreover, about 1/4 of the images in this dataset contain text which are highly correlated to the paired tweets.

\item \textbf{XMedia:} This is the only dataset in the cross-media domain with five different modalities, such as video, audio, image, text, and 3-Dimensional (3D) model. It consists of 20 distinct classes, such as elephant, explosion, bird, dog, etc. Each class contains an overall of 600 media instances: 250 texts, 250 images, 25 videos, 50 audio clips, and 25 3D models. In the dataset's overall collection, different popular websites were used to collect data, i.e., Flickr, YouTube, Wikipedia, 3D Warehouse, and Princeton 3D model search engine. 

\item \textbf{Flickr30K:} the dataset is the extended version of Flickr8k datset \cite{hodosh2013framing}. It consists of 31783 images collected from Flickr. Individual image in this dataset is linked with associated five native English speakers’ descriptive sentences. 

\item \textbf{INRIA-Websearch:} this dataset contains 353 image search queries, along with their meta-data and ground-truth annotations. In total, this dataset consists of 71478 images. 

\item \textbf{IAPR TC-12:} the dataset consists of 20,000 images (plus 20,000 corresponding thumbnails) taken from locations around the world and comprising a varying cross-section of still natural images.The time span used for the collection of images falls within 2001-2005. Moreover, this collection is spatially diverse as the images were collected from more than 30 countries. 

\item \textbf{ALIPR:} the dataset contains annotation results for more than 54,700 images created by users of flickr.com are viewable at the Website: alipr.com.

\item \textbf{LabelMe:} the dataset contains 30,000 images with associated 183 number of labels. The main source of dataset collection was crowd-sourcing through MIT CSAIL Database of objects and scenes\footnote{http://web.mit.edu/torralba/www/database.html}. 

\item \textbf{Corel5K:} the dataset was collected from 50 Corel Stock Photo cds. It consists a total of 5,000 images, with 100 images on the same topic. Individual image is linked with an associated 1-5 keywords with a total of 371 keywords. Before modelling, all the images in the dataset are pre-segmented using normalized cuts \cite{shi2000normalized}. It consists a total of 36 features: 18 color features, 12 texture features and 6 shape features.

\item \textbf{Corel30K:} the dataset is the extended version of previously published dataset called Corel5K. It contains 31,695 images and 5,587 associated words. It exploited 90\% (28,525) and 10\% (3,170) images for training and testing respectively. This dataset is much improved from Corel5K in terms of examples per label and database size, and hence play a significant role in evaluating learning systems. 
\item \textbf{AnnoSearch:} the dataset contains 2.4 million photos collected from popular websites, such as Google\footnote{images.google.com} and the University of Washington (UW)\footnote{http://www.cs.washington.edu/research/imagedatabase/groundtruth/}. The images are of high quality and consists rich associated descriptions, such as title, category and comments from the photographers. Although these descriptions cover to a certain degree the concepts of the associated images. 
\item \textbf{Clickture:} this data set was obtained from the hard work of one-year click log of a commercial image search website. There are 212.3 million triads in this dataset. The triad is mathematically define as:

\begin{equation}
Clickture = \left( {i,k,t} \right),
\end{equation}
A triad $(i, k, t)$ is defined as as image ``$i$'' was clicked ``$t$'' times in the search space of query ``$k$'' in one year by means of different users at different times. The \textit{Clickture} full dataset consists of 40 million unique image and 73.6 million unique text queries. Moreover, this dataset also contains a lite version titled as ``Clickture-Lite'', which consists of 1 million images and 11.7 million text queries. 

\item \textbf{ESP:} the dataset contains more than 10 million images. The key source of dataset collection was crowd-sourcing. The main objective of this cross-media dataset  is to label the most of images over the internet. We envisioned that if our game get a proper gaming site platform, such as Yahoo! Games and allows people to play with interest like other games, it can solve the labeling of most of the images in a time span of weeks. Furthermore, It is predicted that if 5,000 people regularly play this game for 31 days they could assign labels to all Google images. 

\end{enumerate}

\begin{table*}
\begin{center}
\caption{A summary of datasets in cross-media retrieval. For each dataset we identify the modality used to tackle the problem of cross-media retrieval.}
    \label{tab:Training Algorithms}
 \begin{tabular}{| m{3em} |m{1.5cm}| m{2em} | m{3em} | m{5.5cm} | m{3em} |  m{2em} | m{2em} | m{3em}| m{2.2em}|m{2.2em}|}
\hline
 \center{\textbf{Ref}}& \center{\textbf{Dataset}} &  \center{\textbf{Year}} &  \center{\textbf{Data size}} &  \center{\textbf{URL}}&  \center{\textbf{Image }}&  \center{\textbf{Text}} &  \center{\textbf{Tags}}&  \center{\textbf{Video} }& \center{ \textbf{Audio} }&  \textbf{3D Model}\\  
 \hline
 \cite{rasiwasia2010new} & Wikipedia & 2010 &2,866& http://www.svcl.ucsd.edu/projects/crossmodal/ &  \ding{52} & \ding{52} &- &- &-  &-  \\ 
 \hline
\cite{chua2009nus} & Nus Wide & 2009&269,648 &http://lms.comp.nus.edu.sg/research/NUS-WIDE.htm & \ding{52} &  & \ding{52} &- & - &- \\
 \hline
  \cite{hwang2012reading} &Pascal VOC &  2015 &9,963& http://host.robots.ox.ac.uk/pascal/VOC/ & \ding{52} & & \ding{52} & -  &- & -
\\
 \hline
 \cite{young2014image} & Flickr30K & 2014 &31,783& http://shannon.cs.illinois.edu/DenotationGraph/ & \ding{52} & \ding{52}  &- &-  &- &- 
\\
 \hline
\cite{krapac2010improving} & INRIA-Websearch&  2010 &71,478& http://lear.inrialpes.fr/pubs/2010/KAVJ10/ & \ding{52} & \ding{52} &- &- &-  &-
\\
 \hline
\cite{ur2018benchmark} &  FB5K & 2018 &5140& http://ngn.ee.tsinghua.edu.cn/& \ding{52} &-  &- \ding{52}  &- &- &- 
\\
 \hline
  \cite{hu2018twitter100k} & Twitter100K & 2018 &100,000&http://ngnlab.cn/wp-content/uploads/twitter100k.tar & \ding{52} & \ding{52}  &-  &- &- &- 
\\
 \hline
 \cite{peng2018overview} &  Xmedia &  2018 &12,000&http://www.icst.pku.edu.cn/mipl/XMedia & \ding{52} & \ding{52}   & -  &\ding{52} & \ding{52} & \ding{52}
\\
 \hline
 \cite{grubinger2006iapr} & IAPR TC-12&  2006 &20,000& http://imageclef.org/photodata & \ding{52} &\ding{52}  &-  & -  & -& -
\\
 \hline
  \cite{li2011real} & ALPR &  2011 &-& http://alpr.com & \ding{52}  &   &\ding{52} &-  &- &- 
\\
 \hline
 \cite{carneiro2007supervised} & SML &  2007 &-&- &-   &-  &-  &-&- &- 
\\
 \hline
 \ \cite{lavrenko2004model} & Corel5K &  2007 &5000& https://rdrr.io/cran/mldr.datasets/man/corel5k.html&\ding{52}   &  &\ding{52}  &-&- &- 
\\
\hline
\cite{von2004labeling} &  ESP& 2004 &-&- &  \ding{52} &-    & \ding{52} &-&- &- 
\\
\hline
\cite{russell2008labelme} & LabelMe & 2008 &-&  http://www.csail.mit.edu/brussell/research/ LabelMe/intro.html & \ding{52} & - & \ding{52}&-  &- &- 
\\
 \hline
  \cite{wang2006annosearch} & AnnoSearch& 2006 &-&http://wsm.directtaps.net/default.aspx &  \ding{52}  &-   & \ding{52}  &- &- &- 
\\
 \hline
  \cite{hua2013clickage} & Clickture& 2013 &-& http://www.clickture.info & \ding{52}  & \ding{52} &-  &- &- &- 
\\
\hline

\label{tab:dataset}
\end{tabular}
\end{center}

\end{table*}

\section{Challenges in Cross-media Retrieval and Proposed DL based Methods}
In this survey paper, we provide a novel taxonomy according to the challenges faced by multi-modal deep learning approaches in solving cross-media retrieval. In subsection~A, we explain the data representation in cross-modal retrieval because it always difficult task in deep learning. Subsection~B describe the alignment of multimodal. Multimodal alignment is also a challenging task in cross-media retrieval to find the relationship and correlations between different instances in cross modalities. Finally, we also consider the translation in subsection~C that refers to map the data from one modality to another. To tackle the aforementioned problems, we present an extensive review of the state-of-the-art problems and their corresponding solutions to leverage the use of deep learning in cross-media retrieval applications. This new taxonomy will enable researchers to better understand the state-of-the-art problems and solutions, and identify future research directions.

\subsection{Representations}

\begin{figure*}
\centering
\includegraphics[width=17cm,height=6.5cm]{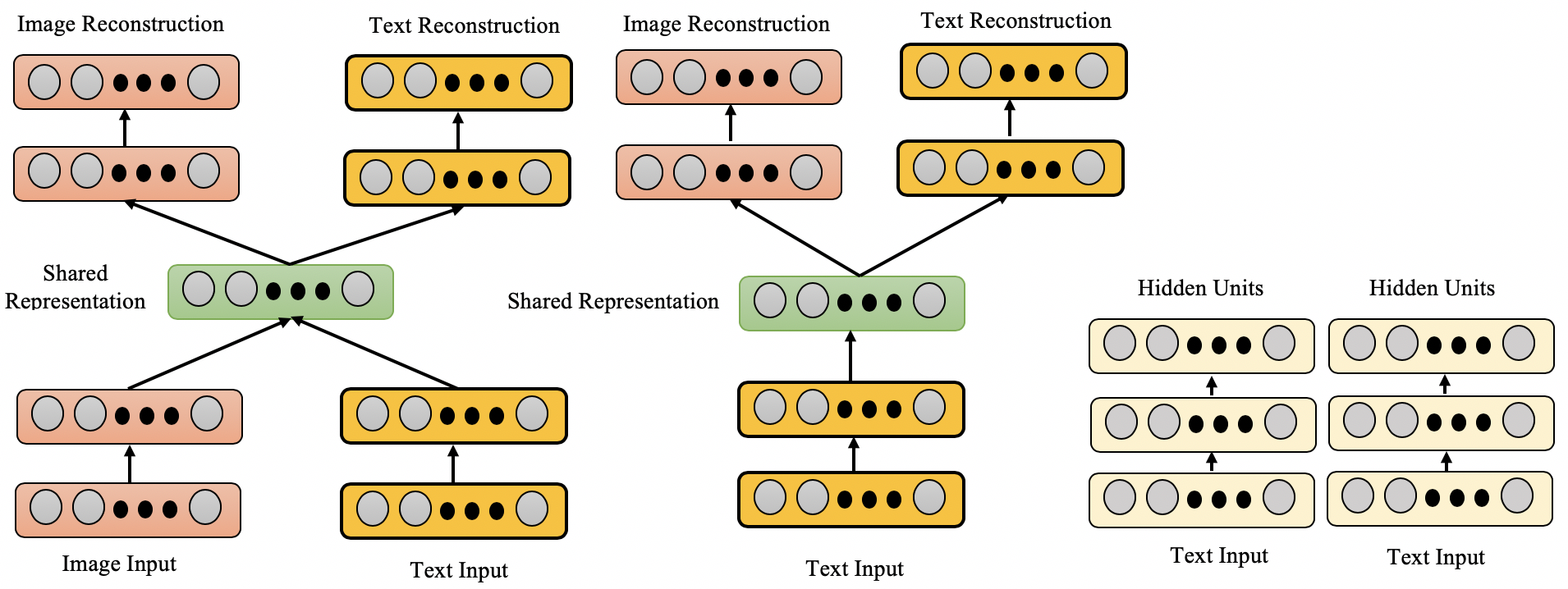}
\caption{An illustration of multimedia for learning shared space representations utilizing deep learning model.}
\label{fig:representation}
\end{figure*}

Data representations in cross-modal retrieval has always been a difficult task in deep learning. Multi-modal representations deal with the representation of data from multiple domains. These representations from different modalities faces several challenges to learn a common semantic space, such as, data concatenation from heterogeneous sources (image, text, video), noise, and missing data handling from various modalities. Semantic data representation tries to learn the correlation across different modalities. Initially, to represent multimodal data in a common semantic space, cross-media correlation learning is performed for feature extraction. Finally, the semantic representations allow the cross-media retrieval to perform search results ranking and summarization. Semantic data representation is mandatory to multi-modal issues, and leverages the performance of any cross-media retrieval model. 

Semantic representations are non-uniform in a low-level feature space. For example, modeling a broad theme, such as \textit{``Asia"}, is more challenging than modeling a specific theme, such as \textit{``sky"}, due to the absence of a significant, unique visual feature that can characterize the concept of \textit{``Asia"}. Therefore, neglecting such semantic representation would be inappropriate. Hence, good representation is indispensable for deep learning models. Bengio \textit{et al.} \cite{bengio2013representation} proposed several ways for good representations - sparsity, smoothness, spatial and temporal coherence etc. It is important to represent data in a meaningful way to enhance the performance of DNN based cross-media retrieval models. 

In a few years, many conventional methods shifted to advanced DNN based methods. For instance, the bag of visual words (BoVW) and scale invariant feature transform (SIFT) were used to represent an image. However, presently CNN \cite{krizhevsky2012imagenet} is used to represent the description of the images. Similarly, Mel-frequency cepstral coefficients (MFCC) have been overcome by deep neural networks in the audio domain for speech recognition \cite{hinton2012deep}.  An overview of
such approaches can be visualized in Fig. \ref{fig:representation}, with representative work summarized in Table \ref{representation}.

\subsubsection{Unsupervised DNNs based Methods}

The major advantage of neural network based joint representations come from their ability to pre-train from unlabeled data when labeled data is not enough for supervised learning. It is also common to fine-tune the resulting representation on a particular task at hand as the representation constructed with unsupervised data is generic and not necessarily optimal for a specific task \cite{wang2015deepp, rehman2018quantum}.
Unsupervised methods used co-occurrence information instead of label information to learn common representations across different modality data. Srivastava \textit{et al.} \cite{srivastava2012learning} learned the representations of multimodal data using Deep Belief Network (DBN). They first model individual media type using a separate DBN model. Then concatenated both networks by learning a mutual RBM at the top.

Chen \textit{et al.} \cite{chen2017deep} proposed conditional generative adversarial networks (CGAN) to achieve cross-modal retrieval of audio-visual generation (e.g, sound and image). Unlike traditional Generative Adversarial Networks (GANs), they make their system to handle cross-modality generation, such as sound to image (S2I) and image to sound (I2S). Furthermore, a fully connected layer and several deconvolution layers of deep convolutional neural networks are used as the image encoder and decoder respectively. Similarly is the case with sound generation. Following the same path, Zhang \textit{et al.} \cite{zhang2017hashgan} proposed a novel adversarial model, called HashGAN. It consists of three main modules: (1) feature learning module for multi-modal data, which uses CNN to extract high level semantic information, (2) generative attention module, which is used to extract foreground and background feature representations, and (3) discriminative hash coding module, which uphold the similarity between cross modalities.

Multi-modal Stacked Auto-Encoders (MSAE) model \cite{wang2014effective} is used to project features from cross-modality into a common latent space for efficient cross-modal retrieval. This model shows significant advantages over current state-of-the-art approaches. First, the non-linear mapping method used in this model is more expressive. Second, since it is an unsupervised learning method, data dependency is minimal. Third, the memory usage is optimized and independent of the training dataset size. Unlike the authors of \cite{fan2017unsupervised}, they proposed an unsupervised deep learning approach in text subspace for cross-media retrieval. They claimed that the proposed text subspace is more efficient and useful as compared to conventional latent subspace.

\subsubsection{Supervised DNNs based Methods}

Ngiam \textit{et al.} \cite{ngiam2011multimodal} were the first to address a multimodal deep learning approach in audio and video retrieval. They trained deep networks for a series of multimodal learning tasks to learn a shared representation between cross modalities and tested it on a single modality, for example, the system was trained with video data but tested with audio data and vice versa.

 Deep Cross-modal Hashing (DCMH) \cite{jiang2016deep} efficiently reveals the correlations among cross modalities. It is an end-to-end learning paradigm, which integrates two parts: (1) feature learning part, and (2) the hash-code learning part. 
Cao \textit{et al.} \cite{cao2016deep} proposed Deep Visual-Semantic Hashing (DVSH) model, which utilized two different DNN models such as CNN and Long Short Term Memory (LSTM) to learn similar representation for visual data and natural language. 
 
Wang \textit{et al.} \cite{wang2015deep} proposed a regularized deep neural network (RE-DNN), which utilized deep CNN features and topic features as visual and textual semantic representation across modalities. This model is able to capture both intra-modal and inter-modal relationships for cross-media retrieval. 
They further improve their work in \cite{wang2016joint, wang2013learning} by concatenating common subspace learning and coupling feature selection into a joint feature learning framework. Unlike previous models, this approach considers both the correlation and feature selection problems at the same time. They learn the projection matrices through linear regression to map cross-modality data into a common subspace, and $\ell_{21}-$norm to select similar/dissimilar features from various feature spaces. Furthermore, the inter-modality and intra-modality similarities are preserved through a multimodal graph regularization.

\subsubsection{Pairwise-based DNNs Methods}

These methods are used to learn a semantic metric distance between cross modalities data for utilizing similar/dissimilar pairs, which is termed as heterogeneous metric learning. 

Social media networks, e.g., Flickr, Facebook, Youtube, Wechat, Twitter, have produced immense data on the web due to which it became the source of high attention. Thus, it plays a significant role in multimedia related applications, including cross-media retrieval. Social media networks are completely different from traditional media network and exhibit unique challenges to data analysis. 1) The data present on social media websites are various and noisy. 2) The data are heterogeneous and present in different modalities, e.g., image, text, video, audio, on the same platform. To predict the link between various instances of social media Yuan \textit{et al.} \cite{yuan2013latent} proposed a brave novel idea on the latent feature learning. To achieve this, they designed a Relational Generative Deep Belief Nets (RGDBN). In this model, they learn the latent feature for social media, which utilized the relationships between social media instances in the network. By integrating the proposed model called the Indian buffet process into the improved DBN, they learn the optimal latent features that best embed both the media content and its relationships. The proposed RGBDBN is able to analyze the correlation between homogeneous and heterogeneous data for cross-media retrieval. 

Following the same path, Wang \textit{et al.} \cite{wang2015image} proposed Modality-Specific Deep
Structure (MSDS) based on modality-specific feature learning. The MSDS model used two different types of CNN to represent raw data in the latent space. The semantic information among the images and texts in the latent space used one-vs more learning scheme. 
Deep Cross-Modal Hashing (DCMH) \cite{jiang2017deep} extends traditional deep
models for cross-modal retrieval, but it can only capture
intra-modal information and ignores inter-modal correlations,
which makes the retrieved results suboptimal. To overcome the aforementioned limitations, a Pairwise Relationship guided Deep Hashing (PRDH) \cite{yang2017pairwise} adopted deep CNN models to learn feature representations and hash codes for individual cross-modality using the end-to-end architecture. Moreover, in this model, the decorrelation constraints are integrated into a single deep architecture to enhance the classification performance of the individual hash bit.

\subsubsection{Rank-based DNNsMethods}

These methods utilize rank lists to learn semantic representations, in which an individual candidate is ranked based on the similarity distance between the query and candidate. In this regard, Hu \textit{et al.} \cite{hu2018twitter100k} achieved the highest efficiency for cross-media retrieval using Dual-CNN's architecture. They used dual CNN to model image and text independently, which is further used to rank the similarity distance between query and candidate. Frome \textit{et al.} \cite{frome2013devise} introduced a novel deep visual-semantic embedding (DeViSE) approach to leverage useful information learned in the text domain, and transfer it to a system trained for visual recognition. Similarly, Weston \textit{et al.} \cite{weston2010large} employed online learning to rank approach, called WSABIE, to train a joint embedding model of labels and images. The authors of \cite{srivastava2012multimodal} developed a Deep Boltzmann Machines (DBM) to represent joint cross-modal probability distribution over sentences and images. Different from RNN-based approaches, Socher \text{et al.} \cite{socher2014grounded} introduced a novel Dependency Tree Recursive Neural Networks (DT-RNNs) model which embed one modality (e.g., sentences) into a vector space using dependency trees in order to retrieve cross-modality (e.g., images). However, these methods reason about the
image only on the global level using a single, fixed-sized representation from the top layer of a
CNN as a description for the entire image. In contrast, the model presented in \cite{karpathy2014deep} clearly elaborated the challenge faced in a complex scene. They formulated a max-margin objective for DNN that learn to embed both image and text into a joint semantic space. The ranking function for joint image-text representations is:

\begin{equation}
{c_G}\left( \theta  \right)\sum\limits_k {\left[ {\begin{array}{*{20}{c}}
{\sum\limits_l {\max \left( {0,{S_{kl}} - {S_{kk}} + \Delta } \right) + } }\\
{\sum\limits_l {\max \left( {0,{S_{lk}} - {S_{kk}} + \Delta } \right)} }
\end{array}} \right]},
\end{equation}
where $\Delta$ is a hyperparameter that we cross-validate. The objective stipulates that the score for true
image-sentence pairs $S_{kk}$ should be higher than $S_{kl}$ or $S_{lk}$ for any $l \ne k$ by at least a margin of $\Delta$.

\begin{table}
\begin{center}
\caption{Summary of DNN based methods for the cross-media representations task.}
    \label{tab:Training Algorithms}
 \begin{tabular}{| m{3cm} ||m{2.5cm}|| m{1.7cm} |}
 \hline 
 Reference & Modalities & Representation \\ [0.5ex] 
 \hline \hline
 \pbox{3cm}{\cite{chen2017deep}, \cite{srivastava2012learning}, \cite{zhang2017hashgan}, \cite{wang2014effective}, \cite{fan2017unsupervised}} &  \pbox{5cm}{Audio and Images \\ Text and Images} & Unsupervised \\ 
 \hline

 \pbox{5cm}{\cite{ngiam2011multimodal}, \cite{jiang2016deep}, \cite{wang2015deep}, \\\cite{wang2016joint, wang2013learning}, \cite{cao2016deep}} & \pbox{3cm}{Audio and Video  \\ Text and Images \\  Images and Audio} & Supervised 
\\
 \hline
 \pbox{5cm} {\cite{yang2017pairwise, yuan2013latent, wang2015image}}  &  \pbox{3cm}{Audio and Images \\ Text and Images} & Pairwise 
\\
 \hline
 \pbox{4.5cm}{\cite{hu2018twitter100k}, \cite{frome2013devise}, \cite{karpathy2014deep}  \cite{weston2010large}, \\ \cite{srivastava2012multimodal}, \cite{socher2014grounded} }&  \pbox{5cm}{Text and Images \\ Label and Images \\ Sentences and Images } & Rank-based
\\
\hline

\end{tabular}
\label{representation}
\end{center}
\label{tab:ref}
\end{table}

\subsection {Alignment}

\begin{figure}
\centering
\includegraphics[width=9cm,height=8cm]{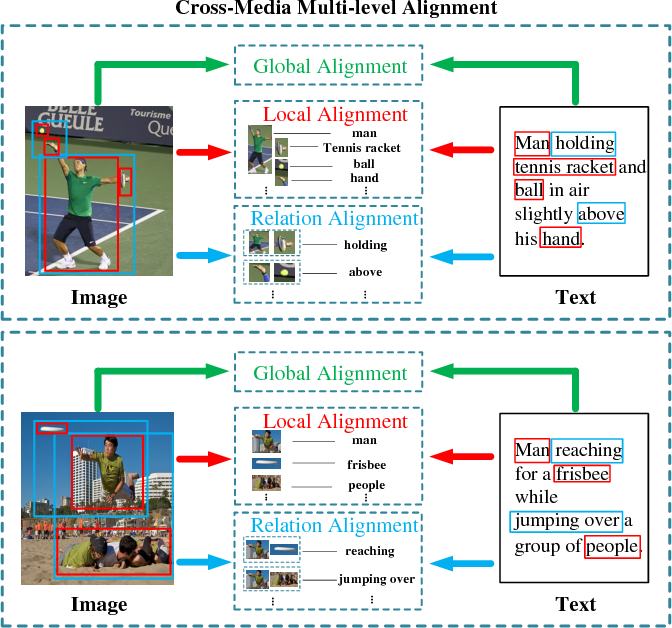}
\caption{An example of cross-media multi-level alignment for correlation learning, which not only explores global alignment between original instances and local alignment between fine-grained patches, but also captures relation alignment lying in the context.}
\label{fig:alignment}
\end{figure}

Multimodal alignment is a challenging task in cross-media retrieval. It basically consists in finding the relationships and correlations between different instances in cross modalities. For example, aligning text and image for a particular website. As the reader get good understanding from both modalities present in a particular webpage rather than just one. Multimodal alignment is significant for cross-media retrieval, as it allows us to retrieve the contents of different modality based on input query (e.g., image retrieval in case of the text as a query, and vice versa) as shown in Fig. \ref{fig:alignment}. Furthermore, we summarized  different DNN based methods for the cross media alignment task in Table \ref{alignment}.

\begin{table}
\begin{center}
\caption{Summary of DNN based methods for the cross-media alignment task.}
    \label{tab:Training Algorithms}
 \begin{tabular}{| m{3cm} ||m{2cm}|| m{2cm} |}
 \hline
 Reference & Modalities & Alignment \\ [0.5ex] 
 \hline\hline
{\cite{rehman2018benchmark, kruskal1983overview, dong2018semi, yan2018joint, chung2018unsupervised} } & \pbox{5cm}{ Image and Text \\ Speech and Text} & Unsupervised \\ 
 \hline
{\cite{qi2018cross} \cite{jourabloo2016large}}  &  \pbox{5cm}{Image and Text \\ Image and gesture} & Supervised \\
 \hline
{\cite{feng2014cross, yuan2018recursive, peng2018ccl, zheng2017dual}} & \pbox{5cm}{Image and Text}& Pairwise 
\\
\hline

\end{tabular}
\label{alignment}
\end{center}
\end{table}

\subsubsection{Unsupervised DNNs based Methods}
Unsupervised methods operate without label information between instances from cross modalities. These methods enforce some constraints on alignment, such as the temporal ordering of sequences and similarity metric existence between the modalities. 

To align multi-view time series Kruskal \textit{et al.} \cite {kruskal1983overview} proposed the Dynamic Time Warping (DTW) approach, which is used to measure the similarity between two instances and find an optimized match between them using time warping (frames insertion). DTW can be used
directly for multimodal alignment by hand-crafting similarity
metrics between modalities; for example Rehman \textit{et al.} \cite{rehman2018benchmark} introduced a novel similarity measurement between texts, images and users' feelings to align images and texts. 

The canonical correlation analysis (CCA) extended the original DTW formulation as it requires a pre-define correlation metric between different modalities \cite{dong2018semi, yan2018joint}. George \textit{et al.} \cite{trigeorgis2018deep} proposed a novel Deep Canonical Time Warping (DCTW) approach to automatically learn composite non-linear representations of multiple time series which are highly correlated and temporally in alignment. Yan \textit{et al.} \cite{yan2015deep} proposed a novel end-to-end approach based on the deep CCA. They formulated the objective function as:

\begin{equation}
\begin{array}{l}
\mathop {\max }\limits_{{k_i},{k_j}} tr\left( {k_i^T\sum {ij{\rm{ }}{k_j}} } \right)\\
s.t.\left[ {k_i^T\sum\nolimits_{ii} {{k_i} = k_j^T\sum\nolimits_{jj} {{k_j} = I} } } \right],
\end{array}
\end{equation}
where \[T = \sum\nolimits_{ii}^{ - 1/2} {\sum\nolimits_{ij} {\sum\nolimits_{jj}^{ - 1/2} {} } },\] and the objective function can be rewritten as follwing.

\begin{equation}
corr\left( {i,j} \right) = tr\left( {{{\left( {{T^T}T} \right)}^{1/2}}} \right).
\end{equation}

Furthermore, Yan \textit{et al.} \cite{yan2015deep} also optimize the memory consumption and speed complexity in the DCCA framework using GPU implementation with CULA libraries, which significantly increase the efficiency as compared to the CPU implementation. 

Chung \textit{et al.} \cite{chung2018unsupervised} proposed an unsupervised cross-modal alignment method to learn the embedding spaces of speech and text. Particularly, the proposed approach used the Speech2Vec \cite{chung2018speech2vec} and Word2Vec \cite{mikolov2013distributed} to learned the respective speech and text embedding spaces. Furthermore, it also attempted to align the two spaces through adversarial training, followed by a refinement method. 

\subsubsection {Supervised DNNs based methods}

Normally, researchers not only focus on the visual regions and keywords, when aligning an image with text, but also between the rely on the correlation between them. Correlation is very important for cross-media learning; however, it is ignored in most of the previous works. For this purpose, Qi \textit{et. al.} \cite{qi2018cross} proposed Cross-media Relation Attention Network
(CRAN) with multi-level alignment. The proposed model was used to efficiently handle the relation between different multimodal domains using multi-level alignment. In another article, Amin \textit{et al.} \cite{jourabloo2016large} proposed a concatenated model of CNN regressor method and a 3-dimensional deep Markov Model (3DMM) to align faces with pose appearance. Dai \textit{et al.} \cite{dai2018cross} proposed a unified framework for cross-media alignment task. They proposed a fused objective function, which contains both CCA-like correlation capability and LDA-like distinguishing capabilities. Further, Jia \textit{et al.} \cite{jia2018deep} proposed an efficient CNN model, which includes three main parts: the visual part is responsible for visual features extraction, the tex part is responsible for text features extraction, and finally the fusion part is responsible to fuse the image and sentences to generate decisive alignment score of the tweet (image and sentence pair).

\subsubsection{Pairwise-based DNNs Mehtods}
With the recent advances of deep learning in multimedia applications, such as image classification \cite{krizhevsky2012imagenet}
and object detection \cite{ren2015faster}, researchers adopt the deep neural network to learn common space for cross-media
retrieval, which aims to fully utilize its considerable ability of modeling a highly nonlinear correlation. Most of the deep learning
based methods construct a multi-pathway network, where each pathway is for the data of one media type. Multiple pathways
are linked at the joint layer to model cross-media correlation. Ngiam et al. propose bimodal autoencoders (Bimodal AE) to
extend the restricted Boltzmann machine (RBM) \cite{ngiam2011multimodal}. They model the correlation by mutual reconstruction
between different media types. Multimodal deep belief network \cite{srivastava2012learning} adopts two kinds
of DBNs to model the distribution over data of different media types, and it constructs a joint RBM to learn cross-media
correlation. Liu \textit{et al.} propose deep canonical correlation analysis (DCCA) to combine traditional CCA with deep network \cite{liu2019new}, which maximizes correlation on the top of two subnetworks. Feng et al. jointly model cross-media
correlation and reconstruction information to perform correspondence autoencoder (Corr-AE) \cite{feng2014cross}.
Furthermore, Yuan \textit{et al.} propose a recursive pyramid network with joint attention (RPJA) \cite{yuan2018recursive}. They construct a hierarchical network structure with stacked learning strategy, which aims to fully exploit both inter-media and intra-media
correlation. Cross-modal correlation learning (CCL) \cite{peng2018ccl} utilizes fine-grained information, and adopts
multi-task learning strategy for better performance. Zheng \textit{et al.} propose a dual-path convolutional network to learn image-text
embedding \cite{zheng2017dual}. They conduct efficient and effective end-to-end learning to directly learn from the
data with the utilization of supervisions. Besides, Plummer \textit{et al.} provide the first large-scale dataset of region-to-phrase correspondences for image description based on Flickr-30K dataset \cite{plummer2015flickr30k}, where image regions depict
the corresponding entities for richer image-to-sentence modelling.

However, the above methods mainly focus on pairwise correlation, which exists in global alignment between original instances of different media types. Although some of they attempt to explore local alignment between fine-grained patches, they all ignore important relation information lying in the context of these fine-grained patches, which can provide rich complementary hints for cross-media correlation
learning. Thus, we propose to fully exploit multi-level cross-media alignment, which can learn the more precise correlation
between different media types.

\subsection{Translation} 

\begin{figure}
\centering
\includegraphics[width=9cm,height=5cm]{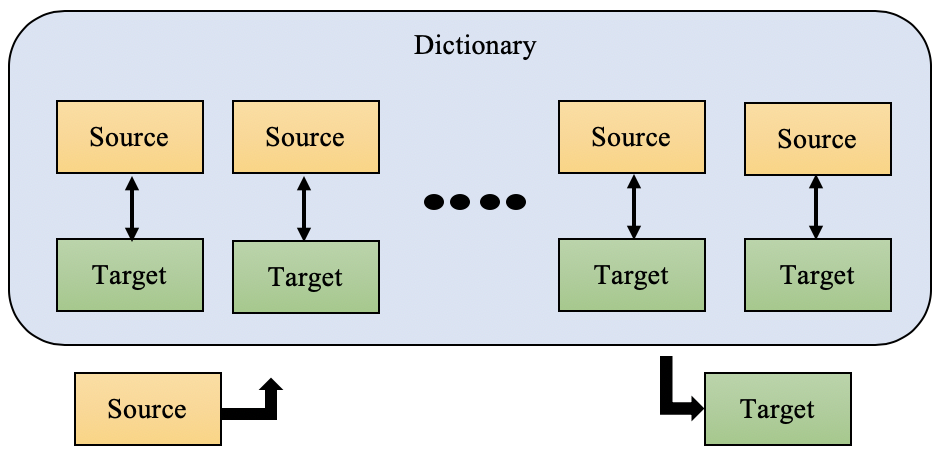}
\vspace{-5mm}
\caption{A generalize description of example-based multimodal translation. It shows that the system retrieves efficient translation as soon as it get a query.}
\label{fig:translation}
\end{figure}

Translation refers to a mapping of data from one modality to another. For example, given a query of one modality, the task is to retrieve different modality of similar information. This task is a critical problem in cross-media retrieval \cite{pereira2014role}, computer vision and multimedia \cite{bernardi2016automatic}. An overview of multi-modal translation can be visualized in Fig. \ref{fig:translation} and the representative work is summarized in Table \ref{tab:translation}.

In recent years, many deep learning based methods have been proposed to elucidate multimodal translation challenges. It is important because the retrieval task from different modalities has to fully understand the visual scene and produce grammatically correct and brief text depicting it. The multimodal translation is a very challenging issue in a deep learning community for several reasons. Foremost, as most of the time, it is hard to choose an appropriate translation for a particular task, where multiple parameters are crucial. Particularly, there is no appropriate correct answer to a query in translation. As there is no common concept of translation to chose which answer is right and which is wrong. 

Another important reason is the variety of media, linguistic, area and culture differences, which further need expertise in the individual domain of translation with image, text and audio channels. We categorize multimodal translation based deep learning methods into two types - \textit{supervised} and \textit{unsupervised}.

\subsubsection{Unsupervised DNNs based Methods}
These approaches normally rely on finding the nearest sample in the dictionary through consensus caption selection and used that as the translated output. Devlin \textit{et al.} \cite{devlin2015language} proposed a \textit{k}-nearest neighbor retrieval approach to achieve translation results.   

In \cite{socher2010connecting} the authors projected words and image regions into a
common space. Moreover, they used unsupervised large text corpora to
learn semantic word representations for cross-media retrieval. Following the same path, Socher \textit{et al.} \cite{socher2013zero} proposed two different deep neural network models for translation. First, they trained a DNN model on many images in order to obtain rich features \cite{coates2011importance}; at the same time, a neural language model \cite{bengio2003neural} was trained to extract embedding representation of text. They further trained a linear mapping between the image features and the text embeddings to decrease the semantic space and link the two modalities. Lample \textit{et al.} \cite{lample2018word} proposed an unsupervised bilingual translation method that can model bilingual dictionary between two different languages. The key benefit of the proposed method is that it does not use any cross-lingual annotated data instead it only uses two monolingual corpora as the source and target language.

\subsubsection{Supervised DNNs based Methods} These approaches rely on label information to retrieve cross-modality instances. Yagcioglu \textit{et al.} \cite{yagcioglu2015distributed} used a CNN-based image representation to translate the
given visual query into a distributional semantics based form. Furthermore, selecting intermediate semantic space for correlation measurement during retrieval is also an alternative way to tackle the problem of translation. Socher \textit{et al.} \cite{socher2014grounded} used intermediate semantic space to translate common representation from text to image and vice versa. Similarly, Xu \textit{et al.} \cite{xu2015jointly} proposed an integrated paradigm that models video and text data simultaneously. Their proposed model contains three fundamental parts: a semantic language model, a video model, and a joint embedding model. The language model was used to embed sentences into a continuous vector space. Whereas in the visual model, DNN was used to capture semantic correlation from videos. Finally, in the fused embedding model, the distance of outputs of the deep video model and language model was minimized in the common space to leverage the semantic correlation between different modality. Cao \textit{et al.} \cite{cao2016deep} proposed a novel Deep Visual-Semantic Hashing (DVSH) model for cross-media retrieval. They generated compact hash codes of visual and text data in a supervised manner, which was able to learn the semantic correlation between image and text data. The proposed architecture fuse joint multimodal embedding and cross-media hashing based on CNN for images, RNN for text and max-margin objective that incorporate both images and text to enable similarity preservation and standard hash codes. Lebret \textit{et al.} \cite{lebret2015phrase} used CNN to generate image representation, which allow the system to infer phrases that describe it. Moreover to predict a set of top-ranked phrases, a trigram constrained language model is proposed to generate syntactically
correct sentences from different subsets phrases. Wei \textit{et al.} \cite{wei2017cross} tackled the cross-media retrieval problem through a novel approach called deep semantic matching (deep-SM). Particularly, images and text are mapped into a joint semantic space using two autonomous DNN models.

The popular benchmark multimodal techniques commonly learns a semantic space for image and text features to find a semantic correlation between them. However, using the same projection into the semantic space for two different tasks such as image-to-text and text-to-image may lead to performance degradation. Therefore, Wei \textit{et al} \cite{wei2016modality} proposed a novel method called Modality-Dependent Cross-media Retrieval (MDCR) to tackle the projection problem into the semantic space efficiently. In their proposed method, they learned two couples of projections for cross-media retrieval despite one couple projection into the semantic space. 
\begin{table}
\begin{center}
\caption{Summary of DNNs based methods for the cross-media translation task.}
    \label{tab:translation}
 \begin{tabular}{| m{3cm} ||m{2cm}|| m{2cm} |}
 \hline
\textbf{ Reference} & \textbf{Modalities} & \textbf{Translation} \\ [0.5ex] 
 \hline \hline
{\cite{socher2010connecting}, \cite{socher2013zero}}  &  Image and Text & Unsupervised \\ 
 \hline
{\cite{yagcioglu2015distributed, socher2014grounded, lebret2015phrase, wei2017cross}  \cite{xu2015jointly}  \cite{cao2016deep} \cite{wei2016modality}}  &  \pbox{5cm}{Image and Text \\ Video and Text \\ Image and Audio \\ Image and Text } & Supervised \\
 \hline

\end{tabular}
\end{center}
\end{table}

\section{Discussion}

\begin{figure}
\center
\includegraphics[width=8cm,height=12cm]{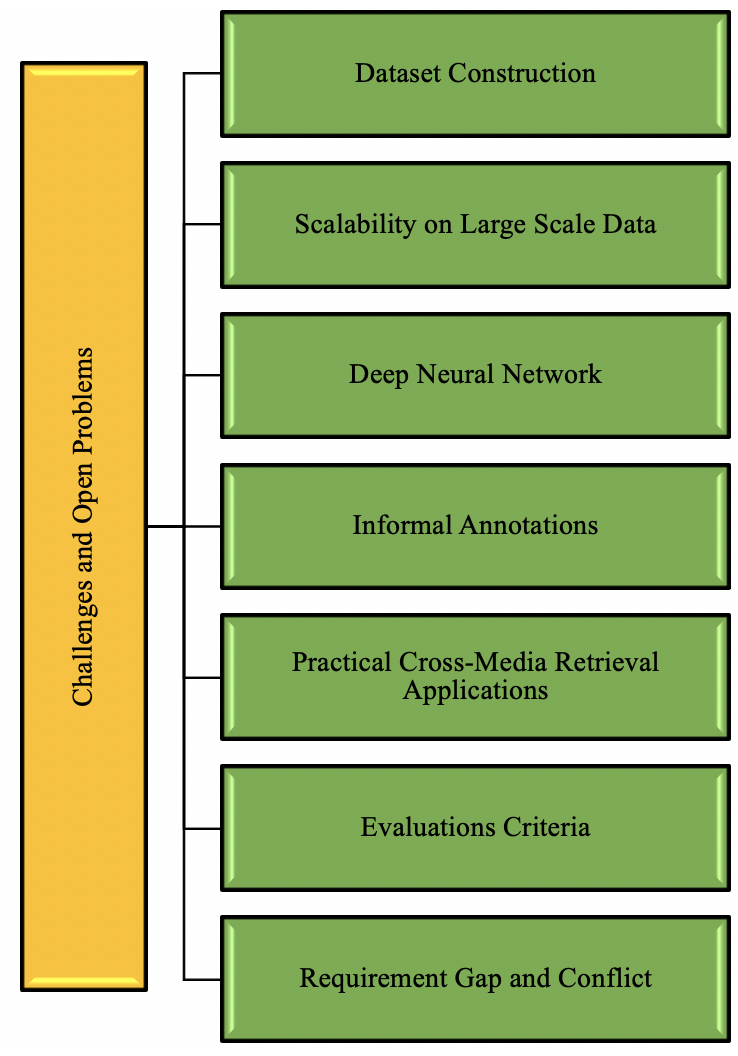}

\caption{Open problems and challenges for future direction}
\label{future-work}
\end{figure}

In this section, we provide a summarized overview of each technical challenge, namely: representation, alignment, and translation, with a discussion of future directions and research problems faced by multi-modal deep learning approaches with application to cross-media retrieval as shown in Fig. \ref{future-work}. We also highlight the lessons and ``best practices" obtained from our review of the existing work.

\subsection{Lessons Learned and Best Practices}
Based on the reviewed papers, we derive a set of lessons learned and ``best practices'' to be considered in implementing and deploying deep learning based cross-media retrieval for solving different challenges, such as representation, alignment, and translation. The key criteria used for solving each challenge is described as follows. 

\subsubsection{Representation}
This section describes four major types of deep learning approaches to solve multimodal representation — unsupervised deep learning, supervised deep learning, pairwise deep learning, and rank based deep learning methods. Unsupervised methods used co-occurrence information instead of label information to learn common representations across different modality data. These methods are commonly used for AVSR, affect, and multimodal gesture recognition. The remaining three representations, project individual modality into a separate space, which often used in applications where single modality is required for retrieval, such as zero-shot learning. Moreover, for the representation task, networks are mostly static. However, in the future, it may be dynamically switching between the modalities \cite{baltruvsaitis2019multimodal, pahuja2019structure}.

\subsubsection{Translation}

Cross-media translation methods are extremely challenging to evaluate. As such, tasks for instance speech recognition have a unique suitable translation, whereas, tasks for instance speech synthesis and image description do not. Most of the time it is hard to choose an appropriate translation for a particular task, where multiple answers are acceptable. However, we can add a number of probabilistic metrics that help in model evaluation. 

Normally, we use the help of human judgment in order to evaluate the aforementioned task. A group of experts has been assigned the task of evaluating individual translation manually through some scale parameter: opinion mining \cite{van2016wavenet, zen2012statistical}, realistic visual speech evaluation \cite{taylor2012dynamic, anderson2013expressive}, media description  \cite{chen2015microsoft, kulkarni2013babytalk, mitchell2012midge, venugopalan2014translating} and correlation and grammatical correctness. On the other hand, preference studies is also an alternate option where various translations are brought forward to the applicant for comparison  \cite{sarfjoo2017using, taylor2017deep}. Though, human judgment is a slow and expensive process. Moreover, they also affected with a different culture, age and gender preferences. It is hoped that by handling the evaluation challenge will be helpful to leverage multimodal translation methods.

\subsubsection{Alignment}
Cross-media alignment has several challenges, which are summarized as follows:
\begin{enumerate}
\item The number of datasets with clearly annotated alignment are scarce. 
\item The development of common similarity metrics between different modalities is hard. 
\item The alignment of different elements in one modality may not have a correspondence in other modality. 
\end{enumerate}

Literature showed that most of the alignment in cross-media focused on the alignment of sequences in an unsupervised manner using graphical models and dynamic programming methods \cite{rehman2019quantum, ur2019learning, rehman2018design}. Most of these methods used hand-crafted similarity measures between different modalities or relied on unsupervised algorithms. However, supervised learning techniques become popular in the current era due to the availability of labeled training data. 

\subsection{Challenges and Open Problems}

\subsubsection{Dataset Construction}

The current state-of-the-art cross-media datasets have significant gaps to fulfil. First, datasets such as Wikipedia dataset\footnote{http://www.svcl.ucsd.edu/projects/crossmodal/} \cite{rasiwasia2010new}, consists of only two different media types i.e., images and texts. In addition to this, Pascal VOC 2012 dataset\footnote{http://host.robots.ox.ac.uk/pascal/VOC/} \cite{hwang2012reading} have only 20 different classes. Although, cross-media concatenate different domains such as images, texts, audio, video and 3D models. Therefore, handling the queries from unknown domain is challenging for the system trained on small dataset  \cite{ur2019unsupervised}. Second, some of the current cross-media datasets are deficient in context information, which results in the decline of cross-media retrieval efficiency. Third, the major limitation in the benchmark cross-media retrieval dataset is the size of the dataset, for instance Xmedia \cite{peng2018overview}, IAPR TC-12 \cite{grubinger2006iapr}, and Wikipedia. This makes the decision challenging for the learning systems due to scarcity of data. Finally, some dataset lacks the proper image labelling aligned with the training set such as, ALIPR \cite{li2011real}, and SML \cite{carneiro2007supervised}. Furthermore, datasets such as ESP \cite{von2004labeling}, LabelMe \cite{russell2008labelme}, and AnnoSearch \cite{wang2006annosearch} withdraw restrictions on the annotation vocabulary, which results in the weak linkage among different modalities semantic gaps. The aforementioned discussion concludes that cross-media retrieval method performance is directly proportional to the nature of the dataset used for evaluation \cite{ur2020optimization}. Therefore, we propose some significant characteristics for a good cross-media retrieval dataset, which are as follow:

\begin{enumerate}
\item Social media platform is the best source for dataset collection as it contains varied domains and informal text language.

\item There must be no constraint in the modality categorization. 

\item Excluding images and texts the dataset also contain other modalities such as video, audio and 3-dimensional (3D) models, which is acceptable in real time scenario.

\item To avoid the overfitting problem during the training of the network. The size of the dataset must be kept significantly large. Also, a large dataset helps the learning algorithm understand the underlying patterns in the data and produce efficient results.

\item The dataset aid in reducing the semantic gap for efficient retrieval by providing coherent visual content descriptors. Also, the datasets with structured alignment between distinct modalities help the learning algorithm to be more robust. 

\end{enumerate}

\subsubsection{Scalability on large-scale data}

With the advancement of technology and the expansion of social media websites around the globe, a large number of multimedia data are produced over the internet. Luckily, deep learning models have exhibit very promising and efficient performance in handling a huge amount of data \cite{najafabadi2015deep} with the help of the Graphical Processing Unit (GPU). Therefore, the need for a scalable and robust model for distributed platforms is significant. Furthermore, it is also noteworthy to investigate further research on effectively organizing individual related modality of data into a common semantic space. We believe compression procedures \cite{serra2017getting} as one of the promising future directions for cross-media retrieval. High-Dimensional input data can be compressed
to compact embedding to reduce the space and computation time during model learning.

\subsubsection{Deep Neural Network}

The work of deep learning on multimodal research is very scarce. Different multimodal hashing techniques are introduced for cross-media retrieval \cite{bronstein2010data, kumar2011learning, zhen2012co, zhen2012probabilistic, song2013inter, wang2014effective, yu2014discriminative, liu2014collaborative, zhang2014large, wu2015quantized, lin2015semantics, long2016composite}. However, these methods are based on shallow architecture, which cannot learn semantic information efficiently between different modalities. Recently, different deep learning models \cite{frome2013devise, kiros2014multimodal, long2015learning,karpathy2015deep, donahue2015long, gao2015you, andreas2015deep, xia2014supervised, lai2015simultaneous, zhu2016deep, cao2016deep} showed that these models were able to extract semantic information between different modalities more efficiently compare to shallow methods. However, they were restricted only to single modality retrieval. One of the promising solutions for the aforementioned problem is transfer learning. It significantly improves the learning task in a specific domain by using knowledge transferred from a different domain. DNN based models are well-matched to transfer learning as it learns both low and high-level features that separate the difference of various cross-media domains.

\subsubsection{Informal annotations}

Social networks websites such as YouTube, Facebook, Instagram, Twitter, and Flickr have produced a large amount of multimodal data over the internet. Generally, this data is poorly organized and has scarce and noisy annotations. However, these annotations provide a correlation between different multimodal data. The key question is how to use the restricted and noisy annotations for a large amount of multimodal data to learn semantic information among the cross-media? 

\subsubsection{Practical Cross-media Retrieval Applications}

As a hot topic these days, practical applications of cross-media retrieval will soon become conceivable due to continuous enhancement in the performance of multimodal efficiency. This will provide easy and flexible retrieval from one modality to another modality. Furthermore, cross-media retrieval is also important in many firms, such as press companies, Television, the entertainment industry, and many others. Currently, people not looking to search for text only but they want to completely visualize things. For example, If you are looking for the installation of a window (operating system) on your machine, it's hard to complete read an article rather than just follow few steps by watching a video. Moreover, the video explains and visualize things better than text and is easily understandable. It is the need for a smart city where people not only search in the same domain but cross-modal searching is also at the fingertips. 

\subsubsection{ Evaluation Criteria} 

In the cross-media community we have seen that each time a model is proposed, it is expected that the model show efficiency against numerous baselines. However, most of the authors did not take it seriously and avail free options for choosing baselines and datasets. This makes several issues in evaluating cross-media models. First, it makes the output prediction score inconsistent. Since individual author reports their own assessed results. By doing this, sometimes, we also encounter conflicts of results. For instance, the original score of the NCF model predicted in its pioneer research work \cite{he2017neural} is ranked very low compared to its variant/modified version \cite{zheng2018mars}. This makes state-of-the-art neural models very difficult. The main question arises here is, how would we solve this issue? Considering other domains such as Natural Language Processing (NLP) or Image Processing they have baseline datasets, such as ImageNet and MNIST for the evaluation of models. Therefore, we strongly believe such a standardized system for the cross-media domain. Second, there must be proper designing of dataset split, particularly, test sets. Without this, in fact, it is challenging to measure the performance of model evaluation. Finally, by using deep learning models it is important to estimate the dataset. As deep learning models performance varies with the amount of data fluctuates. 

\subsubsection{Requirement Gap and Conflict}

Through our review,
we found some blind-spots in DNN-based approaches, such as pairwise based DL methods and rank based DL methods, for solving alignment and translation in cross-media retrieval. The purpose of pairwise based DL methods to
learn common representations through similar/dissimilar
pairs, in which, a semantic metric distance is learned
between data of various modalities, whereas, rank based DL methods are used to learn
common representations for cross-media retrieval through
learning to rank. These approaches are necessary to solve the aforementioned challenges in cross-media retrieval. However, these approaches received little attention in cross-media retrieval and only a few articles have been published in shallow domain \cite{bai2018learning, grangier2018discriminative, mcfee2018metric }. 

Moreover, the deep learning model used by most of the researchers is an individual model for a separate modality. It is strongly recommended that researchers should unfold the recent mathematical theory of deep learning models to investigate the reason why a single model did not achieve benchmark results in cross-media retrieval. It is also encouraged to find out a common semantic space for the features extracted from different modality data using DL models, simultaneously. Furthermore,  the confliction between service quality and retrieval is also noteworthy. For example, DL methods fulfill multiple requirements of feature extraction and distance detection but can be too heavyweight to achieve the real-time constraints of cross-media retrieval. How to strike a balance among contradicting
requirements deserves future studies. The key is to balance feature extraction, similarity measurements, and service quality.

\section{Conclusion}

Multimedia information retrieval is a rapidly growing research field that aims to build models that can validate the information from different modalities. This paper reviewed cross-media retrieval in terms of DNN-based algorithms and presented them in a common classification built upon three technical challenges faced by multimodal researchers: alignment, translation, and representation. For individual challenge, we introduced different sub-classes of DNN-based methods to bridge the media gap, and provide researchers and developers with a better understanding of the underlying problems and the potential solutions of the current deep learning assisted cross-media retrieval research.

\section*{Acknowledgments}
This work was partially supported by The China’s National Key R\&D Program (No.
2018YFB0803600), National Natural Science Foundation of China
(No.61801008), Beijing Natural Science Foundation National
(No. L172049), Scientific Research Common Program
of Beijing Municipal Commission of Education (No.
KM201910005025).

\bibliographystyle{IEEEtran}
\bibliography{references}

\begin{thebibliography}{100}
\providecommand{\url}[1]{#1}
\csname url@samestyle\endcsname
\providecommand{\newblock}{\relax}
\providecommand{\bibinfo}[2]{#2}
\providecommand{\BIBentrySTDinterwordspacing}{\spaceskip=0pt\relax}
\providecommand{\BIBentryALTinterwordstretchfactor}{4}
\providecommand{\BIBentryALTinterwordspacing}{\spaceskip=\fontdimen2\font plus
\BIBentryALTinterwordstretchfactor\fontdimen3\font minus
  \fontdimen4\font\relax}
\providecommand{\BIBforeignlanguage}[2]{{%
\expandafter\ifx\csname l@#1\endcsname\relax
\typeout{** WARNING: IEEEtran.bst: No hyphenation pattern has been}%
\typeout{** loaded for the language `#1'. Using the pattern for}%
\typeout{** the default language instead.}%
\else
\language=\csname l@#1\endcsname
\fi
#2}}
\providecommand{\BIBdecl}{\relax}
\BIBdecl

\bibitem{gasser2019towards}
R.~Gasser, L.~Rossetto, and H.~Schuldt, ``Towards an all-purpose content-based
  multimedia information retrieval system,'' \emph{arXiv preprint
  arXiv:1902.03878}, 2019.

\bibitem{rehman2018benchmark}
S.~U. Rehman, S.~Tu, Y.~Huang, and O.~U. Rehman, ``A benchmark dataset and
  learning high-level semantic embeddings of multimedia for cross-media
  retrieval,'' \emph{IEEE Access}, vol.~6, pp. 67\,176--67\,188, 2018.

\bibitem{peng2018overview}
Y.~Peng, X.~Huang, and Y.~Zhao, ``An overview of cross-media retrieval:
  Concepts, methodologies, benchmarks, and challenges,'' \emph{IEEE
  Transactions on Circuits and Systems for Video Technology}, vol.~28, no.~9,
  pp. 2372--2385, 2018.

\bibitem{dong2018semi}
X.~Dong, J.~Sun, P.~Duan, L.~Meng, Y.~Tan, W.~Wan, H.~Wu, B.~Zhang, and
  H.~Zhang, ``Semi-supervised modality-dependent cross-media retrieval,''
  \emph{Multimedia Tools and Applications}, vol.~77, pp. 3579--3595, 2018.

\bibitem{xia2018cross}
L.~Xia and H.~Zhang, ``Cross—media retrieval via cca—bp neural network,''
  in \emph{IEEE Conference on Industrial Electronics and Applications}, 2018,
  pp. 86--89.

\bibitem{liu2018multi}
R.~Liu, S.~Wei, Y.~Zhao, Z.~Zhu, and J.~Wang, ``Multi-view cross-media hashing
  with semantic consistency,'' \emph{IEEE MultiMedia}, 2018.

\bibitem{shu2018crossfire}
K.~Shu, S.~Wang, J.~Tang, Y.~Wang, and H.~Liu, ``Crossfire: Cross media joint
  friend and item recommendations,'' in \emph{Proceedings of ACM International
  Conference on Web Search and Data Mining}, 2018, pp. 522--530.

\bibitem{xu2018deep}
X.~Xu, L.~He, H.~Lu, L.~Gao, and Y.~Ji, ``Deep adversarial metric learning for
  cross-modal retrieval,'' \emph{World Wide Web}, pp. 1--16, 2018.

\bibitem{hardoon2004canonical}
D.~R. Hardoon, S.~Szedmak, and J.~Shawe-Taylor, ``Canonical correlation
  analysis: An overview with application to learning methods,'' \emph{Neural
  computation}, vol.~16, no.~12, pp. 2639--2664, 2004.

\bibitem{rosipal2005overview}
R.~Rosipal and N.~Kr{\"a}mer, ``Overview and recent advances in partial least
  squares,'' in \emph{International Statistical and Optimization Perspectives
  Workshop" Subspace, Latent Structure and Feature Selection"}.\hskip 1em plus
  0.5em minus 0.4em\relax Springer, 2005, pp. 34--51.

\bibitem{sharma2011bypassing}
A.~Sharma and D.~W. Jacobs, ``Bypassing synthesis: Pls for face recognition
  with pose, low-resolution and sketch,'' \emph{Computer Vision and Pattern
  Recognition (CVPR), IEEE Conference on}, pp. 593--600, 2011.

\bibitem{tenenbaum2000separating}
J.~B. Tenenbaum and W.~T. Freeman, ``Separating style and content with bilinear
  models,'' \emph{Neural Computation}, vol.~12, no.~6, pp. 1247--1283, 2000.

\bibitem{sharma2012generalized}
A.~Sharma, A.~Kumar, H.~Daume, and D.~W. Jacobs, ``Generalized multiview
  analysis: A discriminative latent space,'' in \emph{Computer Vision and
  Pattern Recognition (CVPR), IEEE Conference on}, 2012, pp. 2160--2167.

\bibitem{gong2014multi}
Y.~Gong, Q.~Ke, M.~Isard, and S.~Lazebnik, ``A multi-view embedding space for
  modeling internet images, tags, and their semantics,'' \emph{International
  journal of computer vision}, vol. 106, no.~2, pp. 210--233, 2014.

\bibitem{ranjan2015multi}
V.~Ranjan, N.~Rasiwasia, and C.~Jawahar, ``Multi-label cross-modal retrieval,''
  in \emph{IEEE International Conference on Computer Vision}, 2015, pp.
  4094--4102.

\bibitem{rasiwasia2014cluster}
N.~Rasiwasia, D.~Mahajan, V.~Mahadevan, and G.~Aggarwal, ``Cluster canonical
  correlation analysis,'' in \emph{Artificial Intelligence and Statistics},
  2014, pp. 823--831.

\bibitem{schmidhuber2015deep}
J.~Schmidhuber, ``Deep learning in neural networks: An overview,'' \emph{Neural
  networks}, vol.~61, pp. 85--117, 2015.

\bibitem{lecun2015deep}
Y.~LeCun, Y.~Bengio, and G.~Hinton, ``Deep learning,'' \emph{nature}, vol. 521,
  p. 436, 2015.

\bibitem{liu2017survey}
W.~Liu, Z.~Wang, X.~Liu, N.~Zeng, Y.~Liu, and F.~E. Alsaadi, ``A survey of deep
  neural network architectures and their applications,'' \emph{Neurocomputing},
  vol. 234, pp. 11--26, 2017.

\bibitem{ahmad2019deep}
J.~Ahmad, H.~Farman, and Z.~Jan, ``Deep learning methods and applications,'' in
  \emph{Deep Learning: Convergence to Big Data Analytics}.\hskip 1em plus 0.5em
  minus 0.4em\relax Springer, 2019, pp. 31--42.

\bibitem{deng2014tutorial}
L.~Deng, ``A tutorial survey of architectures, algorithms, and applications for
  deep learning,'' \emph{APSIPA Transactions on Signal and Information
  Processing}, vol.~3, 2014.

\bibitem{pouyanfar2018survey}
S.~Pouyanfar, S.~Sadiq, Y.~Yan, H.~Tian, Y.~Tao, M.~P. Reyes, M.-L. Shyu, S.-C.
  Chen, and S.~Iyengar, ``A survey on deep learning: Algorithms, techniques,
  and applications,'' \emph{ACM Computing Surveys (CSUR)}, vol.~51, no.~5,
  p.~92, 2018.

\bibitem{arulkumaran2017deep}
K.~Arulkumaran, M.~P. Deisenroth, M.~Brundage, and A.~A. Bharath, ``Deep
  reinforcement learning: A brief survey,'' \emph{IEEE Signal Processing
  Magazine}, vol.~34, no.~6, pp. 26--38, 2017.

\bibitem{hussein2017imitation}
A.~Hussein, M.~M. Gaber, E.~Elyan, and C.~Jayne, ``Imitation learning: A survey
  of learning methods,'' \emph{ACM Computing Surveys (CSUR)}, vol.~50, no.~2,
  p.~21, 2017.

\bibitem{chen2014big}
X.-W. Chen and X.~Lin, ``Big data deep learning: challenges and perspectives,''
  \emph{IEEE access}, vol.~2, pp. 514--525, 2014.

\bibitem{najafabadi2015deep}
M.~M. Najafabadi, F.~Villanustre, T.~M. Khoshgoftaar, N.~Seliya, R.~Wald, and
  E.~Muharemagic, ``Deep learning applications and challenges in big data
  analytics,'' \emph{Journal of Big Data}, vol.~2, p.~1, 2015.

\bibitem{hordri2017systematic}
N.~Hordri, A.~Samar, S.~Yuhaniz, and S.~Shamsuddin, ``A systematic literature
  review on features of deep learning in big data analytics,''
  \emph{International Journal of Advances in Soft Computing \& Its
  Applications}, vol.~9, no.~1, 2017.

\bibitem{peng2017overview}
Y.~Peng, X.~Huang, and Y.~Zhao, ``An overview of cross-media retrieval:
  Concepts, methodologies, benchmarks, and challenges,'' \emph{IEEE
  Transactions on circuits and systems for video technology}, vol.~28, no.~9,
  pp. 2372--2385, 2017.

\bibitem{wang2016comprehensive}
K.~Wang, Q.~Yin, W.~Wang, S.~Wu, and L.~Wang, ``A comprehensive survey on
  cross-modal retrieval,'' \emph{arXiv preprint arXiv:1607.06215}, 2016.

\bibitem{liu2010cross}
J.~Liu, C.~Xu, and H.~Lu, ``Cross-media retrieval: state-of-the-art and open
  issues,'' \emph{International Journal of Multimedia Intelligence and
  Security}, vol.~1, no.~1, pp. 33--52, 2010.

\bibitem{tu2017csfl}
S.~Tu, Y.~Huang, G.~Liu \emph{et~al.}, ``Csfl: A novel unsupervised convolution
  neural network approach for visual pattern classification,'' \emph{AI
  Communications}, vol.~30, pp. 311--324, 2017.

\bibitem{benjdira2019unsupervised}
B.~Benjdira, Y.~Bazi, A.~Koubaa, and K.~Ouni, ``Unsupervised domain adaptation
  using generative adversarial networks for semantic segmentation of aerial
  images,'' \emph{Remote Sensing}, vol.~11, p. 1369, 2019.

\bibitem{rehman2016face}
S.~U. Rehman, S.~Tu, Y.~Huang, and Z.~Yang, ``Face recognition: A novel
  un-supervised convolutional neural network method,'' in \emph{IEEE
  International Conference of Online Analysis and Computing Science}, 2016, pp.
  139--144.

\bibitem{yang2018clinical}
Z.~Yang, Y.~Huang, Y.~Jiang, Y.~Sun, Y.-J. Zhang, and P.~Luo, ``Clinical
  assistant diagnosis for electronic medical record based on convolutional
  neural network,'' \emph{Scientific reports}, vol.~8, no.~1, pp. 1--9, 2018.

\bibitem{Rehman2019water}
S.~u. Rehman, Z.~Yang, M.~Shahid, N.~Wei, Y.~Huang, M.~Waqas, S.~Tu
  \emph{et~al.}, ``Water preservation in soan river basin using deep learning
  techniques,'' \emph{arXiv preprint arXiv:1906.10852}, 2019.

\bibitem{donahue2014decaf}
J.~Donahue, Y.~Jia, O.~Vinyals, J.~Hoffman, N.~Zhang, E.~Tzeng, and T.~Darrell,
  ``Decaf: A deep convolutional activation feature for generic visual
  recognition,'' in \emph{International conference on machine learning}, 2014,
  pp. 647--655.

\bibitem{peng2016cross}
Y.~Peng, X.~Huang, and J.~Qi, ``Cross-media shared representation by
  hierarchical learning with multiple deep networks.'' in \emph{IJCAI}, 2016,
  pp. 3846 -- 3853.

\bibitem{yang2017image}
Z.~Yang, Y.-J. Zhang, S.~ur~Rehman, and Y.~Huang, ``Image captioning with
  object detection and localization,'' in \emph{International Conference on
  Image and Graphics}, 2017, pp. 109--118.

\bibitem{vinyals2015show}
O.~Vinyals, A.~Toshev, S.~Bengio, and D.~Erhan, ``Show and tell: A neural image
  caption generator,'' in \emph{Computer Vision and Pattern Recognition (CVPR),
  IEEE Conference on}, 2015, pp. 3156--3164.

\bibitem{ballan2014cross}
L.~Ballan, T.~Uricchio, L.~Seidenari, and A.~Del~Bimbo, ``A cross-media model
  for automatic image annotation,'' in \emph{Proceedings of International
  Conference on Multimedia Retrieval}, 2014, p.~73.

\bibitem{8695043}
C.~{Li}, T.~{Yan}, X.~{Luo}, L.~{Nie}, and X.~{Xu}, ``Supervised robust
  discrete multimodal hashing for cross-media retrieval,'' \emph{IEEE
  Transactions on Multimedia}, pp. 1--1, 2019.

\bibitem{8643797}
J.~{Chi} and Y.~{Peng}, ``Zero-shot cross-media embedding learning with dual
  adversarial distribution network,'' \emph{IEEE Transactions on Circuits and
  Systems for Video Technology}, pp. 1--1, 2019.

\bibitem{8673892}
Y.~{Peng} and J.~{Qi}, ``Reinforced cross-media correlation learning by
  context-aware bidirectional translation,'' \emph{IEEE Transactions on
  Circuits and Systems for Video Technology}, pp. 1--10, 2019.

\bibitem{8691806}
X.~{Huang} and Y.~{Peng}, ``Tpckt: Two-level progressive cross-media knowledge
  transfer,'' \emph{IEEE Transactions on Multimedia}, pp. 1--1, 2019.

\bibitem{8716706}
T.~{Yao}, X.~{Kong}, H.~{Fu}, and Q.~{Tian}, ``Discrete semantic alignment
  hashing for cross-media retrieval,'' \emph{IEEE Transactions on Cybernetics},
  pp. 1--12, 2019.

\bibitem{bengio2013representation}
Y.~Bengio, A.~Courville, and P.~Vincent, ``Representation learning: A review
  and new perspectives,'' \emph{IEEE Transactions on Pattern Analysis and
  Machine Intelligence}, vol.~35, no.~8, pp. 1798--1828, 2013.

\bibitem{bovsnavcki2019deep}
D.~Bo{\v{s}}na{\v{c}}ki, N.~van Riel, and M.~Veta, ``Deep learning with
  convolutional neural networks for histopathology image analysis,'' in
  \emph{Automated Reasoning for Systems Biology and Medicine}.\hskip 1em plus
  0.5em minus 0.4em\relax Springer, 2019, pp. 453--469.

\bibitem{stephenson2019survey}
N.~Stephenson, E.~Shane, J.~Chase, J.~Rowland, D.~Ries, N.~Justice, J.~Zhang,
  L.~Chan, and R.~Cao, ``Survey of machine learning techniques in drug
  discovery,'' \emph{Current drug metabolism}, vol.~20, pp. 185--193, 2019.

\bibitem{bastian2019visual}
B.~T. Bastian, N.~Jaspreeth, S.~K. Ranjith, and C.~Jiji, ``Visual inspection
  and characterization of external corrosion in pipelines using deep neural
  network,'' \emph{NDT \& E International}, pp. 102--134, 2019.

\bibitem{alhnaity2019using}
B.~Alhnaity, S.~Pearson, G.~Leontidis, and S.~Kollias, ``Using deep learning to
  predict plant growth and yield in greenhouse environments,'' \emph{arXiv
  preprint arXiv:1907.00624}, 2019.

\bibitem{cirecsan2012multi}
D.~Cire{\c{s}}an, U.~Meier, and J.~Schmidhuber, ``Multi-column deep neural
  networks for image classification,'' \emph{arXiv preprint arXiv:1202.2745},
  2012.

\bibitem{krizhevsky2012imagenet}
A.~Krizhevsky, I.~Sutskever, and G.~E. Hinton, ``Imagenet classification with
  deep convolutional neural networks,'' in \emph{Advances in Neural Information
  Processing Systems}, 2012, pp. 1097--1105.

\bibitem{marblestone2016toward}
A.~H. Marblestone, G.~Wayne, and K.~P. Kording, ``Toward an integration of deep
  learning and neuroscience,'' \emph{Frontiers in computational neuroscience},
  vol.~10, p.~94, 2016.

\bibitem{AAA12}
R.~Dechter, ``Learning while searching in constraint-satisfaction problems.''
  \emph{University of California, Computer Science Department, Cognitive
  Systems Laboratory}.

\bibitem{AAA14}
F.~J. Gomez and J.~Schmidhuber, ``Co-evolving recurrent neurons learn deep
  memory pomdps,'' in \emph{Proceedings of the 7th annual conference on Genetic
  and evolutionary computation}.\hskip 1em plus 0.5em minus 0.4em\relax ACM,
  2005, pp. 491--498.

\bibitem{AAA15}
I.~Aizenberg, N.~N. Aizenberg, and J.~P. Vandewalle, \emph{Multi-Valued and
  Universal Binary Neurons: Theory, Learning and Applications}.\hskip 1em plus
  0.5em minus 0.4em\relax Springer Science \& Business Media, 2013.

\bibitem{A91}
A.~Ivakhnenko, ``Cybernetic predicting devices,'' Tech. Rep.

\bibitem{A92}
K.~Fukushima, ``Neocognitron: A self-organizing neural network model for a
  mechanism of pattern recognition unaffected by shift in position,''
  \emph{Biological cybernetics}, vol.~36, no.~4, pp. 193--202, 1980.

\bibitem{A93}
S.~Linnainmaa, ``The representation of the cumulative rounding error of an
  algorithm as a taylor expansion of the local rounding errors,''
  \emph{Master's Thesis (in Finnish), Univ. Helsinki}, pp. 6--7, 1970.

\bibitem{A94}
A.~Griewank, ``Who invented the reverse mode of differentiation,''
  \emph{Documenta Mathematica, Extra Volume ISMP}, pp. 389--400, 2012.

\bibitem{A95}
P.~Werbos, ``Beyond regression:" new tools for prediction and analysis in the
  behavioral sciences,'' \emph{Ph. D. dissertation, Harvard University}, 1974.

\bibitem{A96}
P.~J. Werbos, ``Applications of advances in nonlinear sensitivity analysis,''
  in \emph{System modeling and optimization}.\hskip 1em plus 0.5em minus
  0.4em\relax Springer, 1982, pp. 762--770.

\bibitem{A97}
G.~E. Hinton, S.~Osindero, and Y.-W. Teh, ``A fast learning algorithm for deep
  belief nets,'' \emph{Neural computation}, vol.~18, no.~7, pp. 1527--1554,
  2006.

\bibitem{A98}
C.~Poultney, S.~Chopra, Y.~L. Cun \emph{et~al.}, ``Efficient learning of sparse
  representations with an energy-based model,'' in \emph{Advances in neural
  information processing systems}, 2007, pp. 1137--1144.

\bibitem{A99}
Y.~Bengio, Y.~LeCun \emph{et~al.}, ``Scaling learning algorithms towards ai,''
  \emph{Large-scale kernel machines}, vol.~34, no.~5, pp. 1--41, 2007.

\bibitem{A100}
U.~Chavan and D.~Kulkarni, ``Performance issues of parallel, scalable
  convolutional neural networks in deep learning,'' in \emph{Computing,
  Communication and Signal Processing}.\hskip 1em plus 0.5em minus 0.4em\relax
  Springer, 2019, pp. 333--343.

\bibitem{A101}
T.~Chung, B.~Xu, Y.~Liu, C.~Ouyang, S.~Li, and L.~Luo, ``Empirical study on
  character level neural network classifier for chinese text,''
  \emph{Engineering Applications of Artificial Intelligence}, vol.~80, pp.
  1--7, 2019.

\bibitem{A102}
T.~M. Quan, D.~G. Hildebrand, and W.-K. Jeong, ``Fusionnet: A deep fully
  residual convolutional neural network for image segmentation in
  connectomics,'' \emph{arXiv preprint arXiv:1612.05360}, 2016.

\bibitem{A103}
D.~Cire{\c{s}}an, U.~Meier, and J.~Schmidhuber, ``Multi-column deep neural
  networks for image classification,'' \emph{arXiv preprint arXiv:1202.2745},
  2012.

\bibitem{A104}
R.~Parloff, ``Why deep learning is suddenly changing your life,''
  \emph{Fortune. New York: Time Inc}, 2016.

\bibitem{moreira2019contextual}
G.~d. S.~P. Moreira, D.~Jannach, and A.~M. da~Cunha, ``Contextual hybrid
  session-based news recommendation with recurrent neural networks,''
  \emph{arXiv preprint arXiv:1904.10367}, 2019.

\bibitem{zhang2019deeprec}
S.~Zhang, Y.~Tay, L.~Yao, B.~Wu, and A.~Sun, ``Deeprec: An open-source toolkit
  for deep learning based recommendation,'' \emph{arXiv preprint
  arXiv:1905.10536}, 2019.

\bibitem{you2019hierarchical}
J.~You, Y.~Wang, A.~Pal, P.~Eksombatchai, C.~Rosenburg, and J.~Leskovec,
  ``Hierarchical temporal convolutional networks for dynamic recommender
  systems,'' in \emph{The World Wide Web Conference}, 2019, pp. 2236--2246.

\bibitem{zheng2017joint}
L.~Zheng, V.~Noroozi, and P.~S. Yu, ``Joint deep modeling of users and items
  using reviews for recommendation,'' in \emph{Proceedings of the Tenth ACM
  International Conference on Web Search and Data Mining}.\hskip 1em plus 0.5em
  minus 0.4em\relax ACM, 2017, pp. 425--434.

\bibitem{gong2016hashtag}
Y.~Gong and Q.~Zhang, ``Hashtag recommendation using attention-based
  convolutional neural network.'' in \emph{IJCAI}, 2016, pp. 2782--2788.

\bibitem{zhang2017bjoint}
Y.~Zhang, Q.~Ai, X.~Chen, and W.~B. Croft, ``Joint representation learning for
  top-n recommendation with heterogeneous information sources,'' in
  \emph{Proceedings of ACM on Conference on Information and Knowledge
  Management}.\hskip 1em plus 0.5em minus 0.4em\relax ACM, 2017, pp.
  1449--1458.

\bibitem{he2017neural}
X.~He, L.~Liao, H.~Zhang, L.~Nie, X.~Hu, and T.-S. Chua, ``Neural collaborative
  filtering,'' in \emph{Proceedings of the International Conference on World
  Wide Web}, 2017, pp. 173--182.

\bibitem{tay2018latent}
Y.~Tay, L.~Anh~Tuan, and S.~C. Hui, ``Latent relational metric learning via
  memory-based attention for collaborative ranking,'' in \emph{Proceedings of
  the World Wide Web Conference}.\hskip 1em plus 0.5em minus 0.4em\relax
  International World Wide Web Conferences Steering Committee, 2018, pp.
  729--739.

\bibitem{zhang2018neurec}
S.~Zhang, L.~Yao, A.~Sun, S.~Wang, G.~Long, and M.~Dong, ``Neurec: On nonlinear
  transformation for personalized ranking,'' \emph{arXiv preprint
  arXiv:1805.03002}, 2018.

\bibitem{liu2019new}
Y.~Liu, Y.~Li, Y.-H. Yuan, and H.~Zhang, ``A new robust deep canonical
  correlation analysis algorithm for small sample problems,'' \emph{IEEE
  Access}, 2019.

\bibitem{wang2016deep}
W.~Wang, R.~Arora, K.~Livescu, and J.~Bilmes, ``On deep multi-view
  representation learning: objectives and optimization,'' \emph{arXiv preprint
  arXiv:1602.01024}, 2016.

\bibitem{elmadany2016multiview}
N.~E.~D. Elmadany, Y.~He, and L.~Guan, ``Multiview learning via deep
  discriminative canonical correlation analysis,'' in \emph{IEEE International
  Conference on Acoustics, Speech and Signal Processing}, 2016, pp. 2409--2413.

\bibitem{andreas2015deep}
J.~Andreas, M.~Rohrbach, T.~Darrell, and D.~Klein, ``Deep compositional
  question answering with neural module networks. arxiv preprint,'' \emph{arXiv
  preprint arXiv:1511.02799}, vol.~2, 2015.

\bibitem{perera2019deep}
P.~Perera and V.~M. Patel, ``Deep transfer learning for multiple class novelty
  detection,'' in \emph{Proceedings of the IEEE Conference on Computer Vision
  and Pattern Recognition}, 2019, pp. 11\,544--11\,552.

\bibitem{abc}
Y.~Ren and X.~Cheng, ``Review of convolutional neural network optimization and
  training in image processing,'' in \emph{Tenth International Symposium on
  Precision Engineering Measurements and Instrumentation}, vol. 11053.\hskip
  1em plus 0.5em minus 0.4em\relax International Society for Optics and
  Photonics, 2019, pp. 31--36.

\bibitem{Gaurav}
G.~Goswami, N.~Ratha, A.~Agarwal, R.~Singh, and M.~Vatsa, ``Unravelling
  robustness of deep learning based face recognition against adversarial
  attacks,'' in \emph{Thirty-Second AAAI Conference on Artificial
  Intelligence}, 2018.

\bibitem{Wicker2019RobustnessO3}
W.~Matthew and Z.~K. Marta, ``Robustness of 3d deep learning in an adversarial
  setting,'' 2019.

\bibitem{miller1995wordnet}
G.~A. Miller, ``Wordnet: a lexical database for english,'' \emph{Communications
  of the ACM}, vol.~38, pp. 39--41, 1995.

\bibitem{hwang2012reading}
S.~J. Hwang and K.~Grauman, ``Reading between the lines: Object localization
  using implicit cues from image tags,'' \emph{IEEE Transactions on Pattern
  Analysis and Machine Intelligence}, vol.~34, no.~6, pp. 1145--1158, 2012.

\bibitem{ur2018facebook5k}
S.~ur~Rehman, Y.~Huang, S.~Tu, and O.~ur~Rehman, ``Facebook5k: A novel
  evaluation resource dataset for cross-media search,'' in \emph{International
  Conference on Cloud Computing and Security}, 2018, pp. 512--524.

\bibitem{hodosh2013framing}
M.~Hodosh, P.~Young, and J.~Hockenmaier, ``Framing image description as a
  ranking task: Data, models and evaluation metrics,'' \emph{Journal of
  Artificial Intelligence Research}, vol.~47, pp. 853--899, 2013.

\bibitem{shi2000normalized}
J.~Shi and J.~Malik, ``Normalized cuts and image segmentation,'' \emph{IEEE
  Transactions on pattern analysis and machine intelligence}, vol.~22, pp.
  888--905, 2000.

\bibitem{rasiwasia2010new}
N.~Rasiwasia, J.~Costa~Pereira, E.~Coviello, G.~Doyle, G.~R. Lanckriet,
  R.~Levy, and N.~Vasconcelos, ``A new approach to cross-modal multimedia
  retrieval,'' in \emph{Proceedings of the 18th ACM international conference on
  Multimedia}, 2010, pp. 251--260.

\bibitem{chua2009nus}
T.-S. Chua, J.~Tang, R.~Hong, H.~Li, Z.~Luo, and Y.~Zheng, ``Nus-wide: a
  real-world web image database from national university of singapore,'' in
  \emph{Proceedings of the ACM international conference on image and video
  retrieval}.\hskip 1em plus 0.5em minus 0.4em\relax ACM, 2009, p.~48.

\bibitem{young2014image}
P.~Young, A.~Lai, M.~Hodosh, and J.~Hockenmaier, ``From image descriptions to
  visual denotations: New similarity metrics for semantic inference over event
  descriptions,'' \emph{Transactions of the Association for Computational
  Linguistics}, vol.~2, pp. 67--78, 2014.

\bibitem{krapac2010improving}
J.~Krapac, M.~Allan, J.~Verbeek, and F.~Juried, ``Improving web image search
  results using query-relative classifiers,'' in \emph{Computer Vision and
  Pattern Recognition (CVPR), IEEE Conference on}, 2010, pp. 1094--1101.

\bibitem{ur2018benchmark}
S.~ur~Rehman, S.~Tu, Y.~Huang \emph{et~al.}, ``A benchmark dataset and learning
  high-level semantic embeddings of multimedia for cross-media retrieval,''
  \emph{IEEE Access}, 2018.

\bibitem{hu2018twitter100k}
Y.~Hu, L.~Zheng, Y.~Yang, and Y.~Huang, ``Twitter100k: A real-world dataset for
  weakly supervised cross-media retrieval,'' \emph{IEEE Transactions on
  Multimedia}, vol.~20, no.~4, pp. 927--938, 2018.

\bibitem{grubinger2006iapr}
M.~Grubinger, P.~Clough, H.~M{\"u}ller, and T.~Deselaers, ``The iapr tc-12
  benchmark: A new evaluation resource for visual information systems,'' in
  \emph{International workshop ontoImage}, vol.~5, no.~10, 2006.

\bibitem{li2011real}
J.~Li and J.~Z. Wang, ``Real-time computerized annotation of pictures,'' 2011,
  uS Patent 7,941,009.

\bibitem{carneiro2007supervised}
G.~Carneiro, A.~B. Chan, P.~J. Moreno, and N.~Vasconcelos, ``Supervised
  learning of semantic classes for image annotation and retrieval,'' \emph{IEEE
  transactions on pattern analysis and machine intelligence}, vol.~29, no.~3,
  pp. 394--410, 2007.

\bibitem{lavrenko2004model}
V.~Lavrenko, R.~Manmatha, and J.~Jeon, ``A model for learning the semantics of
  pictures,'' in \emph{Advances in neural information processing systems},
  2004, pp. 553--560.

\bibitem{von2004labeling}
L.~Von~Ahn and L.~Dabbish, ``Labeling images with a computer game,'' in
  \emph{Proceedings of the SIGCHI conference on Human factors in computing
  systems}, 2004, pp. 319--326.

\bibitem{russell2008labelme}
B.~C. Russell, A.~Torralba, K.~P. Murphy, and W.~T. Freeman, ``Labelme: a
  database and web-based tool for image annotation,'' \emph{International
  journal of computer vision}, vol.~77, pp. 157--173, 2008.

\bibitem{wang2006annosearch}
X.-J. Wang, L.~Zhang, F.~Jing, and W.-Y. Ma, ``Annosearch: Image
  auto-annotation by search,'' in \emph{Computer Vision and Pattern
  Recognition, IEEE Computer Society Conference on}, vol.~2, 2006, pp.
  1483--1490.

\bibitem{hua2013clickage}
X.-S. Hua, L.~Yang, J.~Wang, J.~Wang, M.~Ye, K.~Wang, Y.~Rui, and J.~Li,
  ``Clickage: Towards bridging semantic and intent gaps via mining click logs
  of search engines,'' in \emph{Proceedings of ACM international conference on
  Multimedia}, 2013, pp. 243--252.

\bibitem{hinton2012deep}
G.~Hinton, L.~Deng, D.~Yu, G.~E. Dahl, A.-r. Mohamed, N.~Jaitly, A.~Senior,
  V.~Vanhoucke, P.~Nguyen, T.~N. Sainath \emph{et~al.}, ``Deep neural networks
  for acoustic modeling in speech recognition: The shared views of four
  research groups,'' \emph{IEEE Signal Processing Magazine}, vol.~29, no.~6,
  pp. 82--97, 2012.

\bibitem{wang2015deepp}
D.~Wang, P.~Cui, M.~Ou, and W.~Zhu, ``Deep multimodal hashing with orthogonal
  regularization.'' in \emph{IJCAI}, vol. 367, 2015, pp. 2291--2297.

\bibitem{rehman2018quantum}
O.~U. Rehman, S.~U. Rehman, S.~Tu, S.~Khan, M.~Waqas, and S.~Yang, ``A quantum
  particle swarm optimization method with fitness selection methodology for
  electromagnetic inverse problems,'' \emph{IEEE Access}, vol.~6, pp.
  63\,155--63\,163, 2018.

\bibitem{srivastava2012learning}
N.~Srivastava and R.~Salakhutdinov, ``Learning representations for multimodal
  data with deep belief nets,'' in \emph{International conference on machine
  learning workshop}, vol.~79, 2012.

\bibitem{chen2017deep}
L.~Chen, S.~Srivastava, Z.~Duan, and C.~Xu, ``Deep cross-modal audio-visual
  generation,'' in \emph{Proceedings of the on Thematic Workshops of ACM
  Multimedia 2017}, 2017, pp. 349--357.

\bibitem{zhang2017hashgan}
X.~Zhang, S.~Zhou, J.~Feng, H.~Lai, B.~Li, Y.~Pan, J.~Yin, and S.~Yan,
  ``Hashgan: Attention-aware deep adversarial hashing for cross modal
  retrieval,'' \emph{arXiv preprint arXiv:1711.09347}, 2017.

\bibitem{wang2014effective}
W.~Wang, B.~C. Ooi, X.~Yang, D.~Zhang, and Y.~Zhuang, ``Effective multi-modal
  retrieval based on stacked auto-encoders,'' \emph{Proceedings of the VLDB
  Endowment}, vol.~7, no.~8, pp. 649--660, 2014.

\bibitem{fan2017unsupervised}
M.~Fan, W.~Wang, P.~Dong, R.~Wang, and G.~Li, ``Unsupervised concept learning
  in text subspace for cross-media retrieval,'' in \emph{Pacific Rim Conference
  on Multimedia}, 2017, pp. 505--514.

\bibitem{ngiam2011multimodal}
J.~Ngiam, A.~Khosla, M.~Kim, J.~Nam, H.~Lee, and A.~Y. Ng, ``Multimodal deep
  learning,'' in \emph{Proceedings of international conference on machine
  learning}, 2011, pp. 689--696.

\bibitem{jiang2016deep}
Q.-Y. Jiang and W.-J. Li, ``Deep cross-modal hashing,'' \emph{Computer Vision
  and Pattern Recognition (CVPR), IEEE Conference on}, 2017.

\bibitem{cao2016deep}
Y.~Cao, M.~Long, J.~Wang, Q.~Yang, and P.~S. Yu, ``Deep visual-semantic hashing
  for cross-modal retrieval,'' in \emph{International Conference on Knowledge
  Discovery and Data Mining}, 2016, pp. 1445--1454.

\bibitem{wang2015deep}
C.~Wang, H.~Yang, and C.~Meinel, ``Deep semantic mapping for cross-modal
  retrieval,'' in \emph{Tools with Artificial Intelligence (ICTAI), IEEE
  International Conference on}, 2015, pp. 234--241.

\bibitem{wang2016joint}
K.~Wang, R.~He, L.~Wang, W.~Wang, and T.~Tan, ``Joint feature selection and
  subspace learning for cross-modal retrieval,'' \emph{IEEE Transactions on
  Pattern Analysis and Machine Intelligence}, vol.~38, no.~10, pp. 2010--2023,
  2016.

\bibitem{wang2013learning}
K.~Wang, R.~He, W.~Wang, L.~Wang, and T.~Tan, ``Learning coupled feature spaces
  for cross-modal matching,'' in \emph{Proceedings of the IEEE International
  Conference on Computer Vision}, 2013, pp. 2088--2095.

\bibitem{yuan2013latent}
Z.~Yuan, J.~Sang, Y.~Liu, and C.~Xu, ``Latent feature learning in social media
  network,'' in \emph{Proceedings of ACM international conference on
  Multimedia}, 2013, pp. 253--262.

\bibitem{wang2015image}
J.~Wang, Y.~He, C.~Kang, S.~Xiang, and C.~Pan, ``Image-text cross-modal
  retrieval via modality-specific feature learning,'' in \emph{Proceedings of
  ACM on International Conference on Multimedia Retrieval}, 2015, pp. 347--354.

\bibitem{jiang2017deep}
Q.-Y. Jiang and W.-J. Li, ``Deep cross-modal hashing,'' in \emph{Computer
  Vision and Pattern Recognition (CVPR), IEEE Conference on}, 2017, pp.
  3232--3240.

\bibitem{yang2017pairwise}
E.~Yang, C.~Deng, W.~Liu, X.~Liu, D.~Tao, and X.~Gao, ``Pairwise relationship
  guided deep hashing for cross-modal retrieval.'' in \emph{Thirty-first AAAI
  conference on artificial intelligence.}, 2017, pp. 1618--1625.

\bibitem{frome2013devise}
A.~Frome, G.~S. Corrado, J.~Shlens, S.~Bengio, J.~Dean, T.~Mikolov
  \emph{et~al.}, ``Devise: A deep visual-semantic embedding model,'' in
  \emph{Advances in neural information processing systems}, 2013, pp.
  2121--2129.

\bibitem{weston2010large}
J.~Weston, S.~Bengio, and N.~Usunier, ``Large scale image annotation: learning
  to rank with joint word-image embeddings,'' \emph{Machine learning}, vol.~81,
  no.~1, pp. 21--35, 2010.

\bibitem{srivastava2012multimodal}
N.~Srivastava and R.~R. Salakhutdinov, ``Multimodal learning with deep
  boltzmann machines,'' in \emph{Advances in neural information processing
  systems}, 2012, pp. 2222--2230.

\bibitem{socher2014grounded}
R.~Socher, A.~Karpathy, Q.~V. Le, C.~D. Manning, and A.~Y. Ng, ``Grounded
  compositional semantics for finding and describing images with sentences,''
  \emph{Transactions of the Association of Computational Linguistics}, vol.~2,
  no.~1, pp. 207--218, 2014.

\bibitem{karpathy2014deep}
A.~Karpathy, A.~Joulin, and L.~F. Fei-Fei, ``Deep fragment embeddings for
  bidirectional image sentence mapping,'' in \emph{Advances in neural
  information processing systems}, 2014, pp. 1889--1897.

\bibitem{kruskal1983overview}
J.~B. Kruskal, ``An overview of sequence comparison: Time warps, string edits,
  and macromolecules,'' \emph{SIAM review}, vol.~25, no.~2, pp. 201--237, 1983.

\bibitem{yan2018joint}
J.~Yan, H.~Zhang, J.~Sun, Q.~Wang, P.~Guo, L.~Meng, W.~Wan, and X.~Dong,
  ``Joint graph regularization based modality-dependent cross-media
  retrieval,'' \emph{Multimedia Tools and Applications}, vol.~77, pp.
  3009--3027, 2018.

\bibitem{chung2018unsupervised}
Y.-A. Chung, W.-H. Weng, S.~Tong, and J.~Glass, ``Unsupervised cross-modal
  alignment of speech and text embedding spaces,'' \emph{arXiv preprint
  arXiv:1805.07467}, 2018.

\bibitem{qi2018cross}
J.~Qi, Y.~Peng, and Y.~Yuan, ``Cross-media multi-level alignment with relation
  attention network,'' \emph{arXiv preprint arXiv:1804.09539}, 2018.

\bibitem{jourabloo2016large}
A.~Jourabloo and X.~Liu, ``Large-pose face alignment via cnn-based dense 3d
  model fitting,'' in \emph{Computer Vision and Pattern Recognition (CVPR),
  IEEE Conference on}, 2016, pp. 4188--4196.

\bibitem{feng2014cross}
F.~Feng, X.~Wang, and R.~Li, ``Cross-modal retrieval with correspondence
  autoencoder,'' in \emph{Proceedings of the ACM international conference on
  Multimedia}.\hskip 1em plus 0.5em minus 0.4em\relax ACM, 2014, pp. 7--16.

\bibitem{yuan2018recursive}
Y.~Yuan and Y.~Peng, ``Recursive pyramid network with joint attention for
  cross-media retrieval,'' in \emph{International Conference on Multimedia
  Modeling}.\hskip 1em plus 0.5em minus 0.4em\relax Springer, 2018, pp.
  405--416.

\bibitem{peng2018ccl}
Y.~Peng, J.~Qi, X.~Huang, and Y.~Yuan, ``Ccl: Cross-modal correlation learning
  with multigrained fusion by hierarchical network,'' \emph{IEEE Transactions
  on Multimedia}, vol.~20, no.~2, pp. 405--420, 2018.

\bibitem{zheng2017dual}
Z.~Zheng, L.~Zheng, M.~Garrett, Y.~Yang, and Y.-D. Shen, ``Dual-path
  convolutional image-text embedding with instance loss,'' \emph{arXiv preprint
  arXiv:1711.05535}, 2017.

\bibitem{trigeorgis2018deep}
G.~Trigeorgis, M.~A. Nicolaou, B.~W. Schuller, and S.~Zafeiriou, ``Deep
  canonical time warping for simultaneous alignment and representation learning
  of sequences,'' \emph{IEEE Transactions on Pattern Analysis and Machine
  Intelligence}, no.~5, pp. 1128--1138, 2018.

\bibitem{yan2015deep}
F.~Yan and K.~Mikolajczyk, ``Deep correlation for matching images and text,''
  in \emph{Computer Vision and Pattern Recognition (CVPR), IEEE Conference on},
  2015, pp. 3441--3450.

\bibitem{chung2018speech2vec}
Y.-A. Chung and J.~Glass, ``Speech2vec: A sequence-to-sequence framework for
  learning word embeddings from speech,'' \emph{arXiv preprint
  arXiv:1803.08976}, 2018.

\bibitem{mikolov2013distributed}
T.~Mikolov, I.~Sutskever, K.~Chen, G.~S. Corrado, and J.~Dean, ``Distributed
  representations of words and phrases and their compositionality,'' in
  \emph{Advances in neural information processing systems}, 2013, pp.
  3111--3119.

\bibitem{dai2018cross}
X.-m. Dai and S.-G. Li, ``Cross-modal deep discriminant analysis,''
  \emph{Neurocomputing}, vol. 314, pp. 437--444, 2018.

\bibitem{jia2018deep}
Y.~Jia, L.~Bai, P.~Wang, J.~Guo, and Y.~Xie, ``Deep convolutional neural
  network for correlating images and sentences,'' in \emph{International
  Conference on Multimedia Modeling}.\hskip 1em plus 0.5em minus 0.4em\relax
  Springer, 2018, pp. 154--165.

\bibitem{ren2015faster}
S.~Ren, K.~He, R.~Girshick, and J.~Sun, ``Faster r-cnn: Towards real-time
  object detection with region proposal networks,'' in \emph{Advances in neural
  information processing systems}, 2015, pp. 91--99.

\bibitem{plummer2015flickr30k}
B.~A. Plummer, L.~Wang, C.~M. Cervantes, J.~C. Caicedo, J.~Hockenmaier, and
  S.~Lazebnik, ``Flickr30k entities: Collecting region-to-phrase
  correspondences for richer image-to-sentence models,'' in \emph{IEEE
  international conference on computer vision}, 2015, pp. 2641--2649.

\bibitem{pereira2014role}
J.~C. Pereira, E.~Coviello, G.~Doyle, N.~Rasiwasia, G.~R. Lanckriet, R.~Levy,
  and N.~Vasconcelos, ``On the role of correlation and abstraction in
  cross-modal multimedia retrieval,'' \emph{IEEE Transactions on Pattern
  Analysis and Machine Intelligence}, vol.~36, no.~3, pp. 521--535, 2014.

\bibitem{bernardi2016automatic}
R.~Bernardi, R.~Cakici, D.~Elliott, A.~Erdem, E.~Erdem, N.~Ikizler-Cinbis,
  F.~Keller, A.~Muscat, and B.~Plank, ``Automatic description generation from
  images: A survey of models, datasets, and evaluation measures,''
  \emph{Journal of Artificial Intelligence Research}, vol.~55, pp. 409--442,
  2016.

\bibitem{devlin2015language}
J.~Devlin, H.~Cheng, H.~Fang, S.~Gupta, L.~Deng, X.~He, G.~Zweig, and
  M.~Mitchell, ``Language models for image captioning: The quirks and what
  works,'' \emph{arXiv preprint arXiv:1505.01809}, 2015.

\bibitem{socher2010connecting}
R.~Socher and L.~Fei-Fei, ``Connecting modalities: Semi-supervised segmentation
  and annotation of images using unaligned text corpora,'' in \emph{Computer
  Vision and Pattern Recognition (CVPR), IEEE Conference on}.\hskip 1em plus
  0.5em minus 0.4em\relax IEEE, 2010, pp. 966--973.

\bibitem{socher2013zero}
R.~Socher, M.~Ganjoo, C.~D. Manning, and A.~Ng, ``Zero-shot learning through
  cross-modal transfer,'' in \emph{Advances in neural information processing
  systems}, 2013, pp. 935--943.

\bibitem{coates2011importance}
A.~Coates and A.~Y. Ng, ``The importance of encoding versus training with
  sparse coding and vector quantization,'' in \emph{Proceedings of
  international conference on machine learning}, 2011, pp. 921--928.

\bibitem{bengio2003neural}
Y.~Bengio, R.~Ducharme, P.~Vincent, and C.~Jauvin, ``A neural probabilistic
  language model,'' \emph{Journal of machine learning research}, vol.~3, pp.
  1137--1155, 2003.

\bibitem{lample2018word}
G.~Lample, A.~Conneau, L.~Denoyer, H.~J{\'e}gou \emph{et~al.}, ``Word
  translation without parallel data,'' 2018.

\bibitem{yagcioglu2015distributed}
S.~Yagcioglu, E.~Erdem, A.~Erdem, and R.~Cakici, ``A distributed representation
  based query expansion approach for image captioning,'' in \emph{Proceedings
  of the 53rd Annual Meeting of the Association for Computational Linguistics
  and the 7th International Joint Conference on Natural Language Processing},
  vol.~2, 2015, pp. 106--111.

\bibitem{xu2015jointly}
R.~Xu, C.~Xiong, W.~Chen, and J.~J. Corso, ``Jointly modeling deep video and
  compositional text to bridge vision and language in a unified framework.'' in
  \emph{In Twenty-Ninth AAAI Conference on Artificial Intelligence.}, vol.~5,
  2015, p.~6.

\bibitem{lebret2015phrase}
R.~Lebret, P.~O. Pinheiro, and R.~Collobert, ``Phrase-based image captioning,''
  \emph{arXiv preprint arXiv:1502.03671}, 2015.

\bibitem{wei2017cross}
Y.~Wei, Y.~Zhao, C.~Lu, S.~Wei, L.~Liu, Z.~Zhu, and S.~Yan, ``Cross-modal
  retrieval with cnn visual features: A new baseline,'' \emph{IEEE Transactions
  on Cybernetics}, vol.~47, no.~2, pp. 449--460, 2017.

\bibitem{wei2016modality}
Y.~Wei, Y.~Zhao, Z.~Zhu, S.~Wei, Y.~Xiao, J.~Feng, and S.~Yan,
  ``Modality-dependent cross-media retrieval,'' \emph{ACM Transactions on
  Intelligent Systems and Technology (TIST)}, vol.~7, p.~57, 2016.

\bibitem{baltruvsaitis2019multimodal}
T.~Baltru{\v{s}}aitis, C.~Ahuja, and L.-P. Morency, ``Multimodal machine
  learning: A survey and taxonomy,'' \emph{IEEE Transactions on Pattern
  Analysis and Machine Intelligence}, vol.~41, pp. 423--443, 2019.

\bibitem{pahuja2019structure}
V.~Pahuja, J.~Fu, S.~Chandar, and C.~J. Pal, ``Structure learning for neural
  module networks,'' \emph{arXiv preprint arXiv:1905.11532}, 2019.

\bibitem{van2016wavenet}
A.~Van Den~Oord, S.~Dieleman, H.~Zen, K.~Simonyan, O.~Vinyals, A.~Graves,
  N.~Kalchbrenner, A.~Senior, and K.~Kavukcuoglu, ``Wavenet: A generative model
  for raw audio,'' \emph{arXiv preprint arXiv:1609.03499}, 2016.

\bibitem{zen2012statistical}
H.~Zen, N.~Braunschweiler, S.~Buchholz, M.~J. Gales, K.~Knill, S.~Krstulovic,
  and J.~Latorre, ``Statistical parametric speech synthesis based on speaker
  and language factorization,'' \emph{IEEE transactions on audio, speech, and
  language processing}, vol.~20, pp. 1713--1724, 2012.

\bibitem{taylor2012dynamic}
S.~L. Taylor, M.~Mahler, B.-J. Theobald, and I.~Matthews, ``Dynamic units of
  visual speech,'' in \emph{Proceedings of the ACM SIGGRAPH/Eurographics
  Symposium on Computer Animation}, 2012, pp. 275--284.

\bibitem{anderson2013expressive}
R.~Anderson, B.~Stenger, V.~Wan, and R.~Cipolla, ``Expressive visual
  text-to-speech using active appearance models,'' in \emph{Computer Vision and
  Pattern Recognition (CVPR), IEEE Conference on}, 2013, pp. 3382--3389.

\bibitem{chen2015microsoft}
X.~Chen, H.~Fang, T.-Y. Lin, R.~Vedantam, S.~Gupta, P.~Doll{\'a}r, and C.~L.
  Zitnick, ``Microsoft coco captions: Data collection and evaluation server,''
  \emph{arXiv preprint arXiv:1504.00325}, 2015.

\bibitem{kulkarni2013babytalk}
G.~Kulkarni, V.~Premraj, V.~Ordonez, S.~Dhar, S.~Li, Y.~Choi, A.~C. Berg, and
  T.~L. Berg, ``Babytalk: Understanding and generating simple image
  descriptions,'' \emph{IEEE Transactions on Pattern Analysis and Machine
  Intelligence}, vol.~35, pp. 2891--2903, 2013.

\bibitem{mitchell2012midge}
M.~Mitchell, X.~Han, J.~Dodge, A.~Mensch, A.~Goyal, A.~Berg, K.~Yamaguchi,
  T.~Berg, K.~Stratos, and H.~Daum{\'e}~III, ``Midge: Generating image
  descriptions from computer vision detections,'' in \emph{Proceedings of
  Conference of the European Chapter of the Association for Computational
  Linguistics}, 2012, pp. 747--756.

\bibitem{venugopalan2014translating}
S.~Venugopalan, H.~Xu, J.~Donahue, M.~Rohrbach, R.~Mooney, and K.~Saenko,
  ``Translating videos to natural language using deep recurrent neural
  networks,'' \emph{arXiv preprint arXiv:1412.4729}, 2014.

\bibitem{sarfjoo2017using}
S.~S. Sarfjoo, C.~Demiro{\u{g}}lu, and S.~King, ``Using eigenvoices and
  nearest-neighbors in hmm-based cross-lingual speaker adaptation with limited
  data,'' \emph{IEEE/ACM Transactions on Audio, Speech, and Language
  Processing}, vol.~25, pp. 839--851, 2017.

\bibitem{taylor2017deep}
S.~Taylor, T.~Kim, Y.~Yue, M.~Mahler, J.~Krahe, A.~G. Rodriguez, J.~Hodgins,
  and I.~Matthews, ``A deep learning approach for generalized speech
  animation,'' \emph{ACM Transactions on Graphics (TOG)}, vol.~36, p.~93, 2017.

\bibitem{rehman2019quantum}
O.~U. Rehman, S.~Yang, S.~Khan, and S.~U. Rehman, ``A quantum particle swarm
  optimizer with enhanced strategy for global optimization of electromagnetic
  devices,'' \emph{IEEE Transactions on Magnetics}, vol.~55, no.~8, pp. 1--4,
  2019.

\bibitem{ur2019learning}
S.~ur~Rehman, Y.~Huang, S.~Tu, and B.~Ahmad, ``Learning a semantic space for
  modeling images, tags and feelings in cross-media search,'' in
  \emph{Pacific-Asia Conference on Knowledge Discovery and Data Mining}.\hskip
  1em plus 0.5em minus 0.4em\relax Springer, 2019, pp. 65--76.

\bibitem{rehman2018design}
O.~U. Rehman, S.~Tu, S.~U. Rehman, S.~Khan, and S.~Yang, ``Design optimization
  of electromagnetic devices using an improved quantum inspired particle swarm
  optimizer.'' \emph{Applied Computational Electromagnetics Society Journal},
  vol.~33, no.~9, 2018.

\bibitem{ur2019unsupervised}
S.~ur~Rehman, S.~Tu, M.~Waqas, Y.~Huang, O.~ur~Rehman, B.~Ahmad, and S.~Ahmad,
  ``Unsupervised pre-trained filter learning approach for efficient convolution
  neural network,'' \emph{Neurocomputing}, vol. 365, pp. 171--190, 2019.

\bibitem{ur2020optimization}
S.~ur~Rehman, S.~Tu, M.~Waqas, O.~Rehman, B.~Ahmad, Z.~Halim, W.~Zhao, and
  Z.~Yang, ``Optimization based training of evolutionary convolution neural
  network for visual classification applications,'' \emph{IET Computer Vision},
  2020.

\bibitem{serra2017getting}
J.~Serr{\`a} and A.~Karatzoglou, ``Getting deep recommenders fit: Bloom
  embeddings for sparse binary input/output networks,'' in \emph{Proceedings of
  the Eleventh ACM Conference on Recommender Systems}.\hskip 1em plus 0.5em
  minus 0.4em\relax ACM, 2017, pp. 279--287.

\bibitem{bronstein2010data}
M.~M. Bronstein, A.~M. Bronstein, F.~Michel, and N.~Paragios, ``Data fusion
  through cross-modality metric learning using similarity-sensitive hashing,''
  in \emph{Computer Vision and Pattern Recognition (CVPR), IEEE Conference on},
  2010, pp. 3594--3601.

\bibitem{kumar2011learning}
S.~Kumar and R.~Udupa, ``Learning hash functions for cross-view similarity
  search,'' in \emph{Twenty-Second International Joint Conference on Artificial
  Intelligence.}, vol.~22, no.~1, 2011.

\bibitem{zhen2012co}
Y.~Zhen and D.-Y. Yeung, ``Co-regularized hashing for multimodal data,'' in
  \emph{Advances in neural information processing systems}, 2012, pp.
  1376--1384.

\bibitem{zhen2012probabilistic}
------, ``A probabilistic model for multimodal hash function learning,'' in
  \emph{Proceedings of the 18th ACM SIGKDD international conference on
  Knowledge discovery and data mining}, 2012, pp. 940--948.

\bibitem{song2013inter}
J.~Song, Y.~Yang, Y.~Yang, Z.~Huang, and H.~T. Shen, ``Inter-media hashing for
  large-scale retrieval from heterogeneous data sources,'' in \emph{Proceedings
  of the ACM SIGMOD International Conference on Management of Data}, 2013, pp.
  785--796.

\bibitem{yu2014discriminative}
Z.~Yu, F.~Wu, Y.~Yang, Q.~Tian, J.~Luo, and Y.~Zhuang, ``Discriminative coupled
  dictionary hashing for fast cross-media retrieval,'' in \emph{Proceedings of
  the international ACM SIGIR conference on Research and development in
  information retrieval}, 2014, pp. 395--404.

\bibitem{liu2014collaborative}
X.~Liu, J.~He, C.~Deng, and B.~Lang, ``Collaborative hashing,'' in
  \emph{Computer Vision and Pattern Recognition (CVPR), IEEE Conference on},
  2014, pp. 2139--2146.

\bibitem{zhang2014large}
D.~Zhang and W.-J. Li, ``Large-scale supervised multimodal hashing with
  semantic correlation maximization.'' in \emph{In Twenty-Eighth AAAI
  Conference on Artificial Intelligence.v}, vol.~1, no.~2, 2014, p.~7.

\bibitem{wu2015quantized}
B.~Wu, Q.~Yang, W.-S. Zheng, Y.~Wang, and J.~Wang, ``Quantized correlation
  hashing for fast cross-modal search.'' in \emph{In Twenty-Fourth
  International Joint Conference on Artificial Intelligence}, 2015, pp.
  3946--3952.

\bibitem{lin2015semantics}
Z.~Lin, G.~Ding, M.~Hu, and J.~Wang, ``Semantics-preserving hashing for
  cross-view retrieval,'' in \emph{Computer Vision and Pattern Recognition
  (CVPR), IEEE Conference on}, 2015, pp. 3864--3872.

\bibitem{long2016composite}
M.~Long, Y.~Cao, J.~Wang, and P.~S. Yu, ``Composite correlation quantization
  for efficient multimodal retrieval,'' in \emph{Proceedings of International
  ACM SIGIR conference on Research and Development in Information Retrieval},
  2016, pp. 579--588.

\bibitem{kiros2014multimodal}
R.~Kiros, R.~Salakhutdinov, and R.~Zemel, ``Multimodal neural language
  models,'' in \emph{International Conference on Machine Learning}, 2014, pp.
  595--603.

\bibitem{long2015learning}
M.~Long, Y.~Cao, J.~Wang, and M.~I. Jordan, ``Learning transferable features
  with deep adaptation networks,'' \emph{arXiv preprint arXiv:1502.02791},
  2015.

\bibitem{karpathy2015deep}
A.~Karpathy and L.~Fei-Fei, ``Deep visual-semantic alignments for generating
  image descriptions,'' in \emph{Computer Vision and Pattern Recognition
  (CVPR), IEEE Conference on}, 2015, pp. 3128--3137.

\bibitem{donahue2015long}
J.~Donahue, L.~Anne~Hendricks, S.~Guadarrama, M.~Rohrbach, S.~Venugopalan,
  K.~Saenko, and T.~Darrell, ``Long-term recurrent convolutional networks for
  visual recognition and description,'' in \emph{Computer Vision and Pattern
  Recognition (CVPR), IEEE Conference on}, 2015, pp. 2625--2634.

\bibitem{gao2015you}
H.~Gao, J.~Mao, J.~Zhou, Z.~Huang, L.~Wang, and W.~Xu, ``Are you talking to a
  machine? dataset and methods for multilingual image question,'' in
  \emph{Advances in neural information processing systems}, 2015, pp.
  2296--2304.

\bibitem{xia2014supervised}
R.~Xia, Y.~Pan, H.~Lai, C.~Liu, and S.~Yan, ``Supervised hashing for image
  retrieval via image representation learning.'' in \emph{Twenty-eighth AAAI
  conference on artificial intelligence}, vol.~1, 2014, p.~2.

\bibitem{lai2015simultaneous}
H.~Lai, Y.~Pan, Y.~Liu, and S.~Yan, ``Simultaneous feature learning and hash
  coding with deep neural networks,'' in \emph{Computer Vision and Pattern
  Recognition (CVPR), IEEE Conference on}, 2015, pp. 3270--3278.

\bibitem{zhu2016deep}
H.~Zhu, M.~Long, J.~Wang, and Y.~Cao, ``Deep hashing network for efficient
  similarity retrieval.'' in \emph{Thirtieth AAAI Conference on Artificial
  Intelligence.}, 2016, pp. 2415--2421.

\bibitem{zheng2018mars}
L.~Zheng, C.-T. Lu, L.~He, S.~Xie, V.~Noroozi, H.~Huang, and P.~S. Yu, ``Mars:
  Memory attention-aware recommender system,'' \emph{arXiv preprint
  arXiv:1805.07037}, 2018.

\bibitem{bai2018learning}
B.~Bai, J.~Weston, D.~Grangier, R.~Collobert, K.~Sadamasa, Y.~Qi, O.~Chapelle,
  and K.~Weinberger, ``Learning to rank with (a lot of) word features,''
  \emph{Information retrieval}, vol.~13, pp. 291--314, 2018.

\bibitem{grangier2018discriminative}
D.~Grangier and S.~Bengio, ``A discriminative kernel-based model to rank images
  from text queries,'' \emph{IEEE Transactions on Pattern Analysis and Machine
  Intelligence}, vol.~10, no. 2018-010, 2018.

\bibitem{mcfee2018metric}
B.~McFee and G.~R. Lanckriet, ``Metric learning to rank,'' in \emph{Proceedings
  of International Conference on Machine Learning}, 2018, pp. 775--782.

\end{thebibliography}

\end{document}